\documentclass[aps,showpacs,twocolumn,amsmath,amssymb,footinbib,natbib]{revtex4-1}
\usepackage{graphicx,epsfig}% Include figure files
\pdfoutput=1
\usepackage{amsfonts}
\usepackage{mathrsfs}
\usepackage{multirow}
\usepackage{color}
\usepackage{graphicx}
\newcommand{\be}{\begin{equation}}
\newcommand{\ee}{\end{equation}}
\newcommand{\bea}{\begin{eqnarray}}
\newcommand{\eea}{\end{eqnarray}}

\begin{document}
%%%%%%%%%%%%%%%%%%%%%%%%%%%%%%%%%%%%%%%%%%%%%%%%%%%%%%%%%%%%%%%%%%%%%
%%%%%%%%%%%%%%%%%%%%%         Title       %%%%%%%%%%%%%%%%%%%%%%%%%%%
%%%%%%%%%%%%%%%%%%%%%%%%%%%%%%%%%%%%%%%%%%%%%%%%%%%%%%%%%%%%%%%%%%%%%

\title{On the (non) existence of superregular boson clouds around extremal Kerr black holes and its connection with number theory}
\date{\today}

%%%%%%%%%%%%%%%%%%%%%%%%%%%%%%%%%%%%%%%%%%%%%%%%%%%%%%%%%%%%%%%%%%%%%
%%%%%%%%%%%%%%%%%%%%     Authors & Addresses  %%%%%%%%%%%%%%%%%%%%%%%
%%%%%%%%%%%%%%%%%%%%%%%%%%%%%%%%%%%%%%%%%%%%%%%%%%%%%%%%%%%%%%%%%%%%%

\author{Gustavo Garc\'ia}
\email{gustavo.garcia@correo.nucleares.unam.mx} 
\author{Marcelo Salgado}
\email{marcelo@nucleares.unam.mx} \affiliation{Instituto de Ciencias
Nucleares, Universidad Nacional Aut\'onoma de M\'exico,\\
 A.P. 70-543, CDMX 04510, M\'exico}

%%%%%%%%%%%%%%%%%%%%%%%%%%%%%%%%%%%%%%%%%%%%%%%%%%%%%%%%%%%%%%%%%%%%%%
%%%%%%%%%%%%%%%%%%%%%           Abstract         %%%%%%%%%%%%%%%%%%%%%
%%%%%%%%%%%%%%%%%%%%%%%%%%%%%%%%%%%%%%%%%%%%%%%%%%%%%%%%%%%%%%%%%%%%%%
\begin{abstract}
We argue about the (non) existence of {\it superregular} scalar clouds (i.e., bound states of a massive and complex-valued scalar field $\Psi$) 
around exact {\it extremal} 
($a = M$) Kerr black holes (BH's) possessing {\it bounded radial derivatives at the horizon} (in Boyer-Lindquist coordinates)
as opposed to similar cloud solutions that exist but with unbounded derivatives in the same coordinate system. The latter solutions have been reported recently both analytically and numerically. 
The superregular clouds cannot be obtained from the regular clouds around subextremal Kerr BH's ($|a|< M$) in the limit of extremality  $(a\rightarrow M)$ 
as in this limit the radial derivatives of $\Psi$ at the horizon $r_H$ diverge when $r_H\rightarrow r_H^{\rm ext}:=M=a$, thus, such superregular clouds must be
analyzed separately. We conclude that the superregular clouds, which 
are found in the {\it exact} extremal scenario ($a = M$), are not continuously connected with the regular ones 
in the limit of extremality  $(a\rightarrow M)$. Remarkably, the spectrum leading to the existence of the radial part of the
full solution of these superregular clouds (which obeys a Teukolsky equation) is given 
by the exact formula $M=a=\frac{1}{2\mu}\sqrt{m^2 + \left[-\kappa+\sqrt{\kappa^2+m^2}\,\right]^2}$, which depends on three (positive) integers: 
the principal number $n$, the {\it magnetic number} $m$, and an integer $j$, related with the {\it type} of regularity at the horizon. 
Here $\kappa= j +n$, and $\mu$ is the mass associated with $\Psi$.
This spectrum depends implicitly on the {\it orbital} number $l$, an integer number that determines 
the existence of well behaved spheroidal harmonics which are associated with the angular part of the cloud solution. Since the separation constants 
that are obtained from the superregularity conditions in the radial part of the solution do {\it not} coincide in general  with the standard separation 
constants required for the spheroidal harmonics to be well behaved on the axis of symmetry, we conclude that non-trivial boson 
clouds having such superregularity conditions cannot exist in the background of an exact extremal Kerr BH.  
The only exception to this conclusion is in the limit $n\rightarrow \infty$ and $m\ll n$. In such a large $n$ limit 
consistency in the separation constants leads to a quadratic Diophantine equation of Pell's type for the integer numbers $(l,m)$. 
Such Pell's equation can be readily solved using standard techniques. In that instance well behaved 
spheroidal harmonics are obtained, and thus, well behaved non-trivial superregular clouds can be computed. 
Of course, this situation, does not preclude the existence of other kind of smooth cloud solutions for any other $n$, not necessarily large 
(e.g. clouds with a non-integer $\kappa$) when 
using a better behaved coordinate system at the horizon (e.g. Wheeler's tortoise coordinate or proper radial distance).
\end{abstract}

 \maketitle

%%%%%%%%%%%%%%%%%%%%%%%%%%%%%%%%%%%%%%%%%%%%%%%%%%%%%%%%%%%%%%%%%%%
%%%%%%%%%%%%%%%%%%%%%%%       Introduction       %%%%%%%%%%%%%%%%%%
%%%%%%%%%%%%%%%%%%%%%%%%%%%%%%%%%%%%%%%%%%%%%%%%%%%%%%%%%%%%%%%%%%%
\section{Introduction}
\label{Introduction}
In a previous work~\cite{Garcia2019} we argued that the (theoretical) existence of
bound states of a massive and complex-valued scalar field $\Psi$ around a Kerr black hole (BH),
dubbed {\it scalar clouds}, and more generally, around axisymmetric and stationary BH that are not necessarily of Kerr type, 
both reported recently by several authors \cite{Herdeiro2014,Herdeiro2015,Hod2012,Hod2013,Hod2017} 
impose important obstructions towards the extension of the no-hair theorems to scenarios with less symmetries than 
the spherically symmetric case. In our previous analysis~\cite{Garcia2019} we restricted ourselves to the nonextremal (i.e. subextremal $|a|< M$) 
and near extremal ($a\approx M$) scenarios in a fixed Kerr background in Boyer-Lindquist coordinates. 
We solved the corresponding eigenvalue problem associated with the complex-valued scalar 
field $\Psi$ and imposed suitable regularity conditions on the field at the 
BH horizon, $r_H$, in addition to other conditions 
that led to the existence of such bound states or clouds, notably, the synchronicity condition $\omega= m\Omega_H$.
In particular, we demanded that radial derivatives of the field were bounded at $r_H$, and thus, that the 
Teukolsky equation associated with the radial part $R(r)$ of the field were satisfied at $r_H$. From these considerations we found 
expressions for the first $R'(r)$ and second $R''(r)$ radial derivatives  
that are always finite at $r_H$ provided the BH was not extremal, i.e., provided $r_H\neq r_H^{\rm ext}:=M=a$, as otherwise, 
those radial derivatives become singular due to the fact that $R'(r_H)$ and $R''(r_H)$ behave like $f(r_H) R(r_H)/(r_H-M)$, 
where $f(r_H)$ represents schematically an expression that is finite at $r_H=M$ in both derivatives, but different in each of them. 
In ~\cite{Garcia2019}, we also argued that despite these divergences when approaching extremality $r_H\rightarrow r_H^{\rm ext}$, 
notably, the divergence in $R'(r_H)$, the resulting scalar clouds with unbounded radial derivatives 
at $r_H^{\rm ext}$ might still make sense physically given that scalar quantities formed from $\Psi$, namely, the ``kinetic term'' $K= |\nabla \Psi|^2= 
g^{ab}(\nabla_a \Psi^*)(\nabla_b \Psi)$, may remain finite at $r_H^{\rm ext}$ due to the presence of a factor $g^{rr}\sim (r-r_H^{\rm ext})^2$ 
associated with the extremal Kerr BH that can compensate the divergence in $R'^2$  when evaluated at 
$r=r_H^{\rm ext}$. This situation indicates that a divergent $R'$ at the horizon in the extremal scenario may be just 
due to a bad coordinate choice and points  perhaps to the necessity of using a better behaved coordinate 
system on the horizon (cf. Sec.~\ref{sec:conclusion}). Nonetheless, in this report we explore the possibility of finding numerically cloud 
solutions with radial gradients that are bounded in the exact extremal case, i.e., when $M=a$, using the Boyer-Lindquist coordinates. 
We dub these clouds {\it superregular clouds} in order to distinguish them from similar clouds that are obtained from the 
subextremal case in the limit $r_H\rightarrow r_H^{\rm ext}=M=a$ and which posses unbounded radial derivatives at $r_H$ in this limit.

To that aim, we analyze the eigenvalue problem for $\Psi$ along the lines presented by us in our previous work~\cite{Garcia2019}, 
but consider {\it ab initio} an exact extremal Kerr background, i.e., we do not consider 
any limit whatsoever from the subextremal case. We then impose regularity on $\Psi$ and its radial derivatives, notably, 
$R'(r)$ and $R''(r)$ at $r_H^{\rm ext}$. We find the exact expressions for $R'(r_H^{\rm ext})$ and $R''(r_H^{\rm ext})$, which are bounded 
and thus, they are not the limit of the corresponding derivatives $R'(r_H)$ and $R''(r_H)$ reported in~\cite{Garcia2019} 
when $r_H\rightarrow r_H^{\rm ext}$ as the latter diverge in this limit, as we stressed above. 
We thus conclude that in the extremal scenario the superregular scalar clouds with finite $R'(r_H^{\rm ext})$ and $R''(r_H^{\rm ext})$, if they actually exist
(i.e. if they can be constructed) are not continuously connected with those obtained from the non-extremal scenario in the extremal limit. 

In the past, Hod~\cite{Hod2012,Hod2013,Hod2017} analyzed the existence of scalar clouds in the extremal and 
near extremal cases and found exact solutions for $\Psi$ with radial gradients that are generically divergent at $r_H^{\rm ext}$, 
except for some ``special'' cases ($\beta=k/2$, where $k$ is an odd positive integer)~\cite{Hod2012,Hod2015}, 
for which those gradients are bounded (see Section~\ref{sec:numerics}). 
Nevertheless, as far as we are aware, in the analyses by Hod, we did not find any further discussion of 
those cases and its connection with the regularity conditions on the scalars formed from $\Psi$. In this paper we make that connection and find that our regularity conditions are related with the resonance condition for $M$ found by Hod ~\cite{Hod2012}. More importantly, by using the
resonance condition together with the regularity conditions we obtain the exact spectra for $M$ given in terms of three integer numbers
that label the cloud solutions: the magnetic number $m$, the principal number $n$ (indicating the number of nodes in the 
the radial function of the solution) and a number $j$ that determines the type of regularity in that function. 
Unfortunately, despite these astonishing features, when looking closer to the angular part of the solution, which is supposed to be given in terms
of the spheroidal harmonics, we realize that the latter do not actually exist, i.e., they do not have an adequate behavior on the axis of symmetry 
at $\theta=0,\pi$, due to the fact that the separation constants, those which
are related to the ``quantization'' conditions, do not coincide with the separation constants that emerge from the regularity conditions 
imposed on the radial part. This lack of consistency on the full solution indicates that non-trivial superregular clouds cannot exist. 
At this respect, we recall that generically, i.e., including the 
subextremal cases, clouds exist only due to the contribution of the rotation of the BH, and thus, due to the contribution of the angular 
part of the solution which is present
only when $m\neq 0$. Thus, if $m=0=l$ automatically one is led to the only possible solution which is the trivial one $\Psi\equiv 0$. 
Apparently the only possibility to conciliate the regularity of the angular and radial parts, and thus, to obtain a consistency 
in the separation constants, is in the limit $n\gg 1$ (i.e. infinite nodes). 
In addition, we found that the pair of integer numbers $(l,m)$ not only have to verify the usual condition $|m|\leq l$, 
but also  a Diophantine equation of Pell's type (see Sections~\ref{sec:betaonehalf} and \ref{sec:betathreehalf}). 
Only then non-trivial superregular boson clouds (of type $\beta=k/2$) seem to exist in the exact extremal Kerr background.

%%%%%%%%%%%%%%%%%%%%%%%%%%%%%%%%%%%%%%%%%%%%%%%%%%%%%%%%%%%%%%%%%%%%%%
%%%%%%%%%%%%%%%%%%%%%%        New Section        %%%%%%%%%%%%%%%%%%%%%
%%%%%%%%%%%%%%%%%%%%%%%%%%%%%%%%%%%%%%%%%%%%%%%%%%%%%%%%%%%%%%%%%%%%%%
\section{The boson clouds}
\label{sec:clouds}
In order to compute the cloud solutions we assume a fixed spacetime background provided by the  
Kerr BH metric in the Boyer-Lindquist coordinates and focus in the extremal case ($M=a$):
\begin{eqnarray}
\label{Kerr} 
\nonumber && ds^2 = -\left(\dfrac{\Delta - a^2\sin^{2}\theta}{\rho^2}\right)dt^2 - \dfrac{2a\sin^2\theta\left(r^2 + a^2 - \Delta\right)}{\rho^2}dtd\varphi \\  
& & + \dfrac{\rho^2}{\Delta}dr^2 + \rho^2d\theta^2 + \left(\dfrac{\left(r^2 + a^2\right)^2 - \Delta a^2\sin^{2}\theta}{\rho^2}\right)\sin^2\theta d\varphi^2, 
\end{eqnarray}
where
\begin{eqnarray}
\Delta &=& r^2 - 2Mr + a^2  = (r - M)^2\;,\\
\rho^2   &=& r^2 + M^2\cos^2\theta \;,
\end{eqnarray}
and $M$ is the {\it mass} and $a$ the {\it angular momentum per mass} associated with the Kerr BH. 
In the subextremal Kerr BH  the horizon is located at largest root $r_+$ of $\Delta(r)= (r-r_+)(r-r_-)$. We denote this
location by $r_H$ and, for the extremal case $M=a$, this location is given by $r_H^{\rm ext}=r_+=r_-=M$.
The angular velocity of the extremal Kerr BH is given by
\begin{equation}
\label{OmH}
\Omega_H^{\rm ext} = \frac{1}{2M}\;.
\end{equation}
We consider a complex-valued massive and free scalar field $\Psi$ that obeys the Klein-Gordon equation 
\begin{equation}
\label{KG}
\Box \Psi= \mu^2 \Psi \;,
\end{equation} 
where $\Box= g^{ab}\nabla_a \nabla_b$ is the covariant d'Alambertian operator, and $g_{ab}$ corresponds to the Kerr metric (\ref{Kerr})
\footnote{The coefficient $\mu$ has dimensions 1/length. The actual coefficient with mass units is given by $\mu_p:= \hbar c \mu/G$,
  and in natural units ($G$ = $c$ = $\hbar$ = 1) both coincide.}.
We are interested in finding ``bound states" solutions or {\it clouds}, so we consider a scalar field $\Psi(t, r, \theta, \varphi)$ 
with temporal and angular dependence of harmonic form, 
\begin{equation}
\label{Psians}
\Psi(t,r,\theta,\varphi)= e^{i (-\omega t + m \varphi)} \phi(r,\theta) \;,
\end{equation}
where $\phi(r,\theta)$ is a real-valued function, $m$ is a non-zero integer (here we focus on $m>0$) and $\omega$ is the frequency of the mode,
which for the bound states is given by the synchronicity condition \cite{Hod2012,Herdeiro2014,Herdeiro2015}:
\begin{equation}
\label{fluxcond2}
\omega = m\Omega_{H}\;,
\end{equation}
where $\Omega_H$ is given by (\ref{OmH}) in the extremal case. The harmonic dependence (\ref{Psians}) is such that the energy-momentum tensor of the
field, as well as its conserved current respect the underlying symmetries of the background spacetime.

%%%%%%%%%%%%%%%%%%%%%%%%%%%%%%%%%%%%%%%%%%%%%%%%%%%%%%%%%%%%%%%%%%%%
%%%%%%%%%%%%%%%%%%%%%%        New Section        %%%%%%%%%%%%%%%%%%%%%
%%%%%%%%%%%%%%%%%%%%%%%%%%%%%%%%%%%%%%%%%%%%%%%%%%%%%%%%%%%%%%%%%%%%%%

The {\it scalar clouds} were analyzed numerically by Herdeiro and
Radu~\cite{Herdeiro2014,Herdeiro2015} in various scenarios and analytically by Hod in the test field
limit~\cite{Hod2012,Hod2013,Hod2017}. In~\cite{Garcia2019} we used an argument in terms of the following integral
\begin{equation}
\label{intident1}
\int_{\cal V} \Big[ \mu^2 \Psi^* \Psi + (\nabla_c \Psi^*) (\nabla^c\Psi)\Big] \sqrt{-g}d^4 x \equiv 0 \;,
\end{equation}
to justify in a simple and heuristic way the existence of the nontrivial boson clouds alluded above. The integral is performed over 
a suitable spacetime volume ${\cal V}$ defined in the region of outer communication of the BH having as boundaries the 
event horizon and two spacelike hypersurfaces extending to spatial infinity where the field $\Psi$ vanishes and the metric becomes the Minkowski metric.

Specifically, in ~\cite{Garcia2019} we showed that in the presence of a {\it subextremal} rotating Kerr BH the above integral can be satisfied even 
when (nontrivial) boson clouds exist (i.e. $\phi(r,\theta)\neq 0$) due to the presence of a non-positive definite contribution 
$m^2  \phi^2 {\cal R}$ in the
{\it kinetic} term: 
\begin{equation}
\label{kinstataxi}
K:= (\nabla_c \Psi^*) (\nabla^c\Psi) 
= m^2  \phi^2 {\cal R} +  g^{IJ}(\nabla_I \phi) (\nabla_J \phi) \;, 
\end{equation}
where
\begin{equation}
\label{haircond}
{\cal R}:= g^{tt}\Omega_H^2 - 2 g^{t\varphi}\Omega_H+ g^{\varphi\varphi} \;,
\end{equation}
and $g^{IJ}(\nabla_I \phi) (\nabla_J \phi)= g^{rr}(\partial_r \phi)^2 + g^{\theta\theta}(\partial_\theta \phi)^2$ is non-negative.
Thus we showed that when boson clouds exist, ${\cal R}\leq 0$ in a spacetime region in order for the term $m^2  \phi^2 {\cal R}$
to compensate the non-negative definite terms $g^{IJ}(\nabla_I \phi) (\nabla_J \phi)$ and $\mu^2  \phi^2$
and thus for the volume integral (\ref{intident1}) to be satisfied~\cite{Garcia2019}. In this report, we show that in the exact
extremal scenario $M=a$, the inequality ${\cal R}\leq 0$ at $\theta=\pi/2$ holds in the domain of outer communication, in contrast with the
subextremal and near extremal cases where ${\cal R}$ is positive near and at $r_H$ and then becomes negative. All this analysis is valid
provided the angular dependence of the field is well behaved, which as we will show, it does not happen in general, but only when $m=0$, in which
case the field disappear, or when $n\gg 1$.

%%%%%%%%%%%%%%%%%%%%%%%%%%%%%%%%%%%%%%%%%%%%%%%%%%%%%%%%%%%%%%%%%%%%%%
%%%%%%%%%%%%%%% Section: Numerical Results  %%%%%%%%%%%%%%%%%%%%%%%%%%
%%%%%%%%%%%%%%%%%%%%%%%%%%%%%%%%%%%%%%%%%%%%%%%%%%%%%%%%%%%%%%%%%%%%%%
\section{The Teukolsky Equation, Regularity Conditions and Numerical results}
\label{sec:numerics}
The Klein-Gordon Eq.~(\ref{KG}) is solved using separation of variables with the mode expansion in the following form
\begin{equation}
\Psi_{nlm}\left(t, r, \theta, \varphi\right) = R_{nlm}\left(r\right)S_{lm}\left(\theta\right)e^{im\varphi}e^{-i\omega t}\;,
\end{equation}
where we have introduced explicitly the labels $(n,l,m)$ that are associated with each solution given this set of integer numbers.

The angular functions $S_{lm}(\theta)$ are the spheroidal harmonics (SH) which obey the angular equation
\begin{widetext}
\begin{equation}
 \label{angularE}
\frac{1}{\sin\theta}\frac{d}{d\theta}\left(\sin\theta\frac{dS_{lm}}{d\theta}\right) + \left(K_{lm} + M^2(\mu^2 - \omega^2)\sin^2\theta - \frac{m^2}{\sin^2\theta}\right)S_{lm} = 0 \;,
\end{equation}
\end{widetext}
where $K_{lm}$ are separations constants ($|m|\leq l$) 
that couple the radial and angular parts of the Klein-Gordon equation (\ref{KG}):
\begin{eqnarray}
\label{Klmsubext}
K_{lm} &=& l(l + 1) - a^2(\mu^2 - \omega^2) + \sum_{k=1}^{\infty}c_{k}a^{2k}(\mu^2 - \omega^2)^k \nonumber \\
      &=& l(l + 1) - \lambda + \sum_{k=1}^{\infty}c_{k}\lambda^{k} \;.
\end{eqnarray}
where 
\begin{equation}
\lambda= a^2(\mu^2 - \omega^2) \;.
\end{equation}
The separation constants $K_{lm}$ ensure that the SH are well behaved on the axis of symmetry. 

The expansion coefficients $c_k$ are provided in Ref.~\cite{Abramowitz}. 
Several remarks on these separation 
constants are in order. Even if the SH have been analyzed thoroughly in the past (see \cite{Flammer1957,Press1992,Li98,Hod2015b,Zhao2017,Stein2019}), there is not a consensus on how to compute 
the coefficients $c_k$, and there is also a lack of a deeper mathematical analysis on the convergent properties of the above 
series. Empirical (numerical) evidence shows that when $|\lambda|\leq 1$, 
the separation constants can be approximated by ${\cal K}_{lm}= l(l+1)$ (i.e. as if there was no rotation) and 
the SH become well behaved except near the axis of symmetry ($\theta=0,\pi$). When including the series 
as in (\ref{Klmsubext}) (usually with the first three terms is sufficient) 
the separation constants only change by a small amount, but the behavior of the 
SH at $\theta=0,\pi$ improves. For instance, in the subextremal cases, and typically when $|m|\leq 3$, 
$\lambda$ is bounded $0<\lambda\leq 1$ (e.g. for $a>0$), and as mentioned, with three terms in the series one obtains a good approximation 
for $K_{lm}$: the relative change in the separation constants $K_{lm}$ compared with ${\cal K}_{lm}$
is $\approx 1.5\%$ when $m=1$ and $l=1$, and for larger $m$ and $l$ the change is even smaller. That is, the series contributes slightly 
as $m$ increases. Alternatively, one can compute $K_{lm}$ with a better accuracy than using the series by solving numerically Eq.~(\ref{angularE}) 
with a shooting method. On the other hand, when $|\lambda| > 1$ a much better precision for $K_{lm}$ is required for 
the SH to be well behaved on the axis of symmetry, and thus, more terms in the series are needed. This situation is encountered 
for clouds with larger $m$. In order to appreciate much better such situation, we note that 
an adequate asymptotic behavior for the radial function (see below) demands $\omega/\mu <1$. 
Thus, even if $\omega$ is bounded by $\mu$ one can have $a/\mu>1$, which is precisely the coefficient that appears in $\lambda$. 
Specifically, for $m=3$, and as the BH approaches extremality one 
finds $a/\mu>1$ (cf. Table III of Ref. \cite{Garcia2019}).

In the exact extremal scenario we will face a much more serious problem. When 
imposing the regularity conditions on the derivatives of the radial functions a completely different kind of 
separation constants will emerge (cf. Eqs.~(\ref{Klm}) and (\ref{Klm2}) below) and thus consistency between these two kind 
of separation constants, and thus, consistency in the full cloud solution, is in jeopardy, except in the large $n$ limit.

The radial functions $R_{nlm}$ obey the radial Teukolsky equation~\cite{Teukolsky1972}
\begin{widetext}
\begin{equation}
\label{radialE}
\Delta\frac{d}{dr}\left(\Delta\frac{dR_{nlm}}{dr}\right) + \left[\mathcal{H}^2 + \left(2mM\omega - K_{lm} - \mu^2\left(r^2 + M^2\right)\right)\Delta\right]R_{nlm} = 0
\;, 
\end{equation}
\end{widetext}
where 
\begin{equation}
\label{H}
\mathcal{H}:=\left(r^2 + M^2\right)\omega - Mm = \frac{m}{2M}(r-M)(r+M) \;.
\end{equation}
The last equality arises from the condition (\ref{fluxcond2}) and from Eq.~(\ref{OmH}). Thus we appreciate that 
$\mathcal{H}$ vanishes precisely at the BH horizon $r=M=r_{H}^{\rm ext}$. 

In order to find physically meaningful configurations that represent {\it bound states} it is necessary to impose regularity conditions of the scalar field 
$\Psi(t, r, \theta, \varphi)$ at BH horizon. We thus consider that the field and its derivatives are bounded at the horizon in order for the scalars 
constructed from $\Psi$ are finite there. In particular, we impose $R_{nlm}(r)$, $R'_{nlm}(r)$ and 
$R''_{nlm}(r)$ to be bounded at $r = r_{H}^{\rm ext}$, where {\it primes} indicate derivatives with respect to the radial coordinate.
These regularity conditions seem reasonable and sufficient to compute well behaved clouds solutions, but are not necessary in 
this coordinate system. In fact, 
one of the aims of this paper is precisely to show that these conditions are extremely restrictive and leads generically to 
inconsistencies. We will elaborate more about this issue in the last section and a propose an alternative treatment to deal with 
the extremal scenario numerically using a more appropriate radial coordinate.

In our previous work~\cite{Garcia2019} we obtained such regularity conditions for the subextremal case and found that the expressions for 
$R'_{nlm}(r_H)$ and $R''_{nlm}(r_H)$ were finite except when $r_H= r_{H}^{\rm ext}=M$. Since in the extremal scenario it is difficult (if not impossible) 
to implement exactly those two conditions in a computer given that the ``initial data'' (data at $r_H= r_{H}^{\rm ext}$) 
diverge, we were able to approach the extremal case only in the limit $r_H\rightarrow r_{H}^{\rm ext}$, which corresponds to the {\it near extremal case}, 
and reported several results in plots and tables. Nonetheless, Hod~\cite{Hod2012} found exact cloud solutions for the exact extremal scenario with
regular radial functions $R$ (we omit for the moment the labels $n,l,m$ in $R$) at $ r_{H}^{\rm ext}$
given by\footnote{We remarked a typo in the original paper \cite{Hod2012}: the dimensionless radial coordinate $z$ is defined by  
$z= 2M(r/r_{H}^{\rm ext} -1)\sqrt{\mu^2-\frac{m^2}{4}}$, and thus, a factor $1/M^2$ is missing within the square root. This factor is important in order 
  for the numerical solutions to match the actual exact solution. The actual expression should read as Eq.(\ref{zeta})
  which is dimensionless.}
\begin{eqnarray}
\label{RHod}
  R &=& A z^{-\frac{1}{2}+\beta} e^{-\frac{1}{2}z} {\cal M}(\frac{1}{2}+\beta-\kappa,1+2\beta,z) \;,\\
\label{betaHod}
\beta^2 &=&  K_{lm}+\frac{1}{4}- 2m^2 + 2M^2\mu^2 \;, \\
\label{kappaHod}
\kappa &=& \frac{m^2 - 2M^2\mu^2}{\sqrt{4M^2\mu^2- m^2}} \;,\\
\label{zeta}
z  &=& (\frac{r}{M} -1)\sqrt{4M^2\mu^2-m^2} \;,
\end{eqnarray}
where ${\cal M}(a,b,z)$ is the confluent hypergeometric function, and $A$ is a {\it normalization} constant. According to \cite{Hod2012}, 
the regularity condition on $R$ at the horizon requires 
$\Re(\beta) \geq 1/2$ (hereafter we consider only a real-valued $\beta$). Now, the {\it generic} case $\beta > 1/2$ leads to solutions where $R$ is regular, but its 
derivatives may blow up at the horizon. In particular, if $1/2< \beta < 3/2$, $R'$ and $R''$ diverge at the horizon. For 
values of $\beta$ in this range the cloud solutions seem to correspond to the extremal limit obtained from the subextremal case that we computed recently when exploring numerically the radial solutions~\cite{Garcia2019}. However, it is not exactly so because in ~\cite{Garcia2019} 
we assumed $R_H=1$ and solution (\ref{RHod}) has $R_H=0$. Thus, from our subextremal solutions as we approach extremality the derivatives 
blow up, but the condition $R_H=1$ remains fixed. On the other hand, we cannot start computations from the subextremal clouds with 
$R_H=0$, and then try to approach the extremal limit, as in this case 
the regularity conditions lead to $R'_H=0=R''_H$~\cite{Garcia2019}, and thus the trivial (identically vanishing) cloud solution is recovered. 
One can in principle recover Hod's extremal solution numerically for this range of $\beta$ but by starting the numerical computation at 
some $r$ different from $r_H$, while keeping the original radial coordinate (cf. [18]), although, as we discuss at the end, 
it is perhaps better to use a different radial coordinate. Instead, in this paper we try to investigate if its possible to find cloud solutions with bounded radial derivatives at the horizon. After all, this is what it has been done for the subextremal case with successful results.
 
In this regard we appreciate from (\ref{RHod}) that for values
 $\beta= k/2$ where $k$ is an odd positive integer (i.e. $\beta= 1/2,3/2,5/2$, etc) one can find different kind of non-trivial 
radial solutions where  $R$ seems to be smooth at the horizon (i.e. where $R$ and all its derivatives are bounded). 
This is what we call {\it superregular clouds}. Depending on the specific value for $k$ several derivatives 
of the radial functions may vanish at the horizon, not only $R_H$ itself.

In order to recover the superregular cloud scenario associated with the values $\beta=k/2$ following our numerical approach, we impose regularity 
conditions at the horizon using Eq.(\ref{radialE}). One can differentiate Eq.(\ref{radialE}) and from the resulting equation it is easy to see that 
when assuming $|R'(r_{H}^{\rm ext})|<\infty $, $|R''(r_{H}^{\rm ext})|<\infty$, $|R'''(r_{H}^{\rm ext})|<\infty$, and $R(r_{H}^{\rm ext})$ bounded as well, 
one is lead to the following condition 
\begin{equation}
\label{CNEg1}
R'_{nlm}(r_{H}^{\rm ext}) = \frac{\left(2\mu^2M^2 - m^2\right) R_{nlm}(r_{H}^{\rm ext}) }{M[2(1+m^2) - K_{lm}- 2\mu^2M^2]}\;.
\end{equation} 
The exact value for $R''(r_{H}^{\rm ext})$ is found by differentiating Eq.(\ref{radialE}) two times assuming 
$|R''''(r_{H}^{\rm ext})|<\infty$:
\begin{widetext}
\begin{equation}
\label{CNEg2} 
R''_{nlm}(r_{H}^{\rm ext}) = -\frac{\left(m^2 - 4\mu^2M^2\right) R_{nlm}(r_{H}^{\rm ext}) + 4M \left(m^2 - 2\mu^2M^2\right) R'_{nlm}(r_{H}^{\rm ext})}
{2M^2 [ 2(3+m^2) - K_{lm}- 2\mu^2M^2]} \;.
\end{equation}
\end{widetext}
The cloud solutions to be obtained using these new conditions, notably those with $R_{nlm}(r_{H}^{\rm ext})\neq 0$, 
will not be connected continuously with the extremal solutions found from the 
subextremal ones in the limit $r_H\rightarrow M$ as in this limit $R'(r_H)$ and $R''(r_H)$ blow up (cf. Eqs. (77) and (78) of \cite{Garcia2019}), 
while in the current situation the radial derivatives (\ref{CNEg1}) and (\ref{CNEg2}) are bounded.

At this point we have two possibilities leading to superregular clouds. One is taking 
$R_{nlm}(r_{H}^{\rm ext})\neq 0$, which as we analyze below, corresponds to $\beta=1/2$, and the other one is 
$R_{nlm}(r_{H}^{\rm ext})= 0$ with $R'_{nlm}(r_{H}^{\rm ext})\neq 0$ and $R''_{nlm}(r_{H}^{\rm ext})\neq 0$, 
which is associated with $\beta=3/2$. The latter situation is possible if we choose the separation constants judiciously, 
as otherwise, for $R_{nlm}(r_{H}^{\rm ext})= 0$ without any restriction on the values for the separation constants 
the radial derivatives (\ref{CNEg1}) and (\ref{CNEg2}) vanish as well and so the radial function vanishes everywhere. 
This is clearly a not very interesting scenario. In the following two sections we consider for simplicity only the two possibilities 
$\beta=1/2$ and $\beta=3/2$. For larger values of $\beta=k/2$ the same conclusions hold albeit the details are different (see the discussion at the end of 
Sec.~\ref{sec:betathreehalf}).

\subsection{The superregular extremal scenario  $R_{nlm}(r_{H}^{\rm ext})\neq 0$ ($\beta=1/2$) and first Pell-Diophantine equation} 
\label{sec:betaonehalf}
In this scenario $R_{nlm}(r_{H}^{\rm ext})$ is finite although arbitrary and 
Eqs. (\ref{CNEg1}) and (\ref{CNEg2}) fixes the values $R'_{nlm}(r_{H}^{\rm ext})$ and $R''_{nlm}(r_{H}^{\rm ext})$ when 
$R_{nlm}(r_{H}^{\rm ext})$ is provided. The full solution is then obtained when the spectrum is found 
(i.e. the adequate value for $M$ leading to localized solutions or bound states).
On the other hand,  when Eq.~(\ref{radialE}) is evaluated at the horizon with the condition $R_{nlm}(r_{H}^{\rm ext})\neq 0$ and again assuming 
$|R'(r_{H}^{\rm ext})|<\infty $, $|R''(r_{H}^{\rm ext})|<\infty$ the separation constants become
\begin{equation}
\label{Klm}
K_{m,\frac{1}{2}}^{\rm ext} = 2(m^2 - \mu^2M^2) \;.
\end{equation}
These separation constants are rather different from (\ref{Klmsubext}), which are the ones needed to ensure regularity of the SH on the axis of symmetry. 
In the subextremal case this situation does not happen since the regularity conditions on the radial function do not impose further constraints on 
(\ref{Klmsubext}).
 
Notice, however, that (\ref{Klm}) can be obtained from Eq.~(\ref{betaHod}) for $\beta=1/2$, this is why we introduced an index `$\frac{1}{2}$' 
in (\ref{Klm}). Using the separation constants (\ref{Klm}) in the regularity conditions (\ref{CNEg1}) and (\ref{CNEg2}) 
we find respectively
\begin{widetext}
\begin{eqnarray}
\label{CNE1}
R'_{nlm}(r_{H}^{\rm ext}) &=& \frac{\left[2\mu^2M^2 - m^2\right] }{2M}R_{nlm}(r_{H}^{\rm ext})\;, \\
\label{CNE2}
R''_{nlm}(r_{H}^{\rm ext}) &= & \frac{\left(4\mu^2M^2 -m^2\right)+ 2\left(m^2 - 2\mu^2M^2\right)^2}{12M^2}R_{nlm}(r_{H}^{\rm ext})\;. 
\end{eqnarray}
\end{widetext}
In order to compute specific radial solutions we choose $R_{nlm}(r_{H}^{\rm ext})=1$ for simplicity, as we did before in the 
subextremal scenario~\cite{Garcia2019}, which leads to nonvanishing $R'_{nlm}(r_{H}^{\rm ext})$ and $R''_{nlm}(r_{H}^{\rm ext})$.

As for the case $R_{nlm}(r_{H}^{\rm ext})= 0$, it is also possible to obtain non-vanishing radial derivatives at the horizon, by 
considering separation constants different from (\ref{Klm}). Such regularity conditions leads also to non-trivial radial functions 
and are associated with the value $\beta=3/2$. In the next subsection we tackle this scenario.
 
\begin{figure}[h!]
\centering
\includegraphics[width=0.5\textwidth]{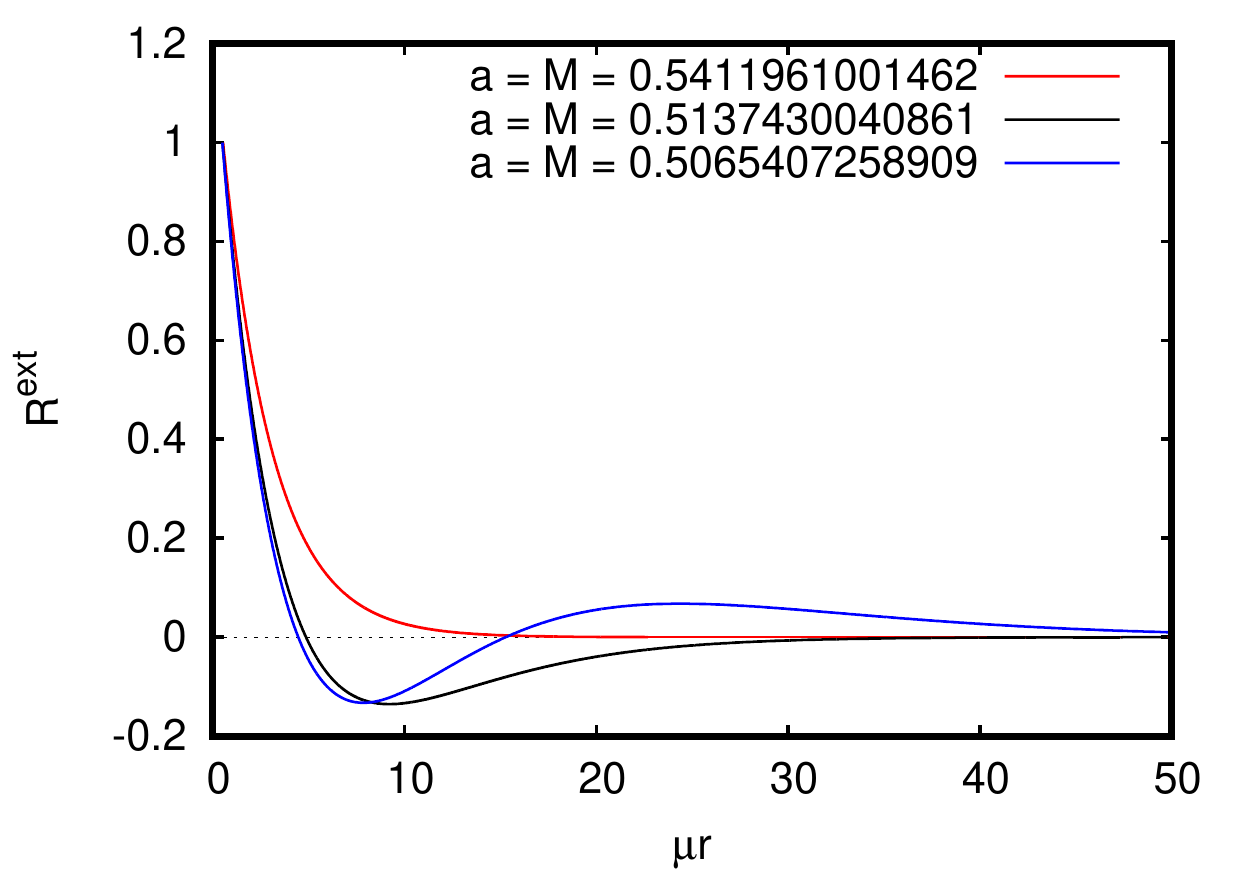}
\caption{Radial solutions $R^{\rm ext}$ with magnetic number $m=1$, and principal numbers 
$n = 0,1,2$ (number of nodes) in extremal Kerr backgrounds with the horizon located at $r_H^{\rm ext} = M$. 
The corresponding eigenvalues $a= M$ are displayed (in units $1/\mu$). Notice that the solutions and its radial 
gradient $R'$ are bounded at $r=r_H^{\rm ext}=M$.}
\label{fig:R11}
\end{figure}

 From Eq.~(\ref{CNE1}) we appreciate that the coefficient $\left[2\mu^2M^2 - m^2\right]=-(m-\sqrt{2}\mu M)(m+\sqrt{2}\mu M)$ 
is negative if $m> \sqrt{2}\mu M$. We are interesting in $R'_{nlm}(r_{H}^{\rm ext})<0$ because this is what systematically results from the numerical analysis 
(cf. Fig.~\ref{fig:R11}). Moreover, from this analysis we also corroborate that $R''_{nlm}(r_{H}^{\rm ext}) >0$, so together with 
the condition $m> \sqrt{2}\mu M$, we conclude from Eq.~(\ref{CNE2}) that a 
sufficient condition for $R''_{nlm}(r_{H}^{\rm ext}) >0$ is $m< 2\mu M$. These conditions translate into the band
\begin{equation}
\label{Mband}
\frac{m}{2} < \mu M < \frac{m}{\sqrt{2}} \;.
\end{equation}

 According to Hod~\cite{Hod2012} this is precisely the band that restricts the spectra for $M$, but here 
this band is found from empirical considerations using the regularity conditions and the numerical analysis. In Hod's exact analysis, 
the resonance condition for the existence of bound states reads: 
\begin{equation}
  \label{kappasmooth0}
  \kappa= \frac{1}{2} + \beta + n \;,
  \end{equation}
implying $\kappa>0$ since $\beta\geq 1/2$. Using the resonance condition (\ref{kappasmooth0}) and Eq.~(\ref{kappaHod}) Hod found the 
band (\ref{Mband}) for $M$ and then the spectra for some values of the parameters. Later, we will
discuss some additional features associated with the band (\ref{Mband}) and the large $n,m$ limits.

Finally, by imposing the boundedness of the radial function (\ref{RHod}) 
at the horizon, and using the resonance condition (\ref{kappasmooth0}) Hod obtained from Eq.~(\ref{RHod}) the expression~\cite{Hod2012}
  \begin{equation}
    R=A z^{-\frac{1}{2}+\beta} e^{-\frac{1}{2}z} L_n^{(2\beta)}(z) \;,
    \label{RHod2}
    \end{equation}
where $L_n^{(2\beta)}(z)$ are the generalized Laguerre polynomials. This is the radial function that is regular at the 
horizon ($z=0)$, but with unbounded gradients for generic values $\beta> 1/2$. The radial function with bounded gradients 
at the horizon are associated with the ``special'' half-integer values $\beta=k/2$ that we mentioned before.

In this section we restrict first to the value $\beta=1/2$ and compare our 
numerical solution with the analytic one provided by Hod\footnote{\label{foot:unbound} For $\beta> 1/2$ the radial function 
vanishes at the horizon ($z=0$) and if in addition $\beta \neq k/2$ the radial function has unbounded gradients. 
In our previous analysis~\cite{Garcia2019} we considered the near extremal 
case with unbounded gradients at the horizon, but assuming $R_H=1$. Therefore in order to compare with Hod's analytic solution (\ref{RHod2}) 
departing from the subextremal configurations and then approach the extremal scenario 
we would require to set $R_H=0$. However, by doing so one obtains $R'_{H}=0=R''_{H}$ and so our numerical solution becomes 
the trivial one $R(r)\equiv 0$. So from the numerical point of view it is rather ``tricky'' to recover the non-trivial solution that 
approaches the extremal analytic solution when imposing $R_H=0$. In order to overcome this drawback, we imposed instead conditions 
associated with the maximum 
of (\ref{RHod2}) at $z_{\rm max}= 2\beta-1$ (for simplicity we considered the nodless case $n=0$ and $m=l=1$) 
and then integrated numerically towards and away the horizon, and checked that 
we indeed recovered numerically and with a good approximation Hod's solution (\ref{RHod2}) with $r_H=0.525/\mu\approx M\approx a$. 
While this remark is relevant for those interested in recovering numerically those kind of solutions, it is not, however, important for the 
problem at hand since here we are mainly concerned in finding clouds with bounded gradients at the horizon}. 
In that case (\ref{kappasmooth0}) reduces to
\begin{equation}
  \label{kappasmooth}
  \kappa= 1 + n \;,
  \end{equation}
and the radial function has $R_H\neq 0$.

Let us recall that we use the ``quantum numbers" $(n, l, m)$ 
to label the possible solutions, where the non-negative integer $n$ determines the number of nodes of the radial function $R_{nlm}$. Here $l$ is also 
a positive integer, and the ``magnetic" number $m$ is also an integer satisfying $|m| \leq l$ (for simplicity we consider only $m>0$).

In order to proceed in our quest for numerical cloud solutions we require to find the spectra $M=a=r_H$ for a given $(n,l,m)$ leading to bound states. In 
particular, localized radial functions. 
In the subextremal scenario that we analyzed previously~\cite{Garcia2019}, we fixed $r_H$ in advance and used $a$ as a shooting parameter leading to 
radial solutions that vanished asymptotically. From this value $a$ it was straightforward to obtain $M$, $\Omega_H$ and the other values of the
parameters associated with the Kerr BH. Since in the {\it extremal} scenario $a = M=r_{H}^{\rm ext}$ 
the location of the horizon $r_{H}^{\rm ext}$ cannot be fixed in advance, but becomes the shooting parameter as well.

According to the separation constants (\ref{Klm}), for the extremal scenario 
we need only to fix the number $m$, and using a shooting method we are able to find the values $a=M$
that lead to radial solutions that vanish asymptotically and that have the desired number of nodes $n$. We do this by using the
  exact values (\ref{Klm}) for the separation constants which do {\it not} depend explicitly on the number $l$. Due to this feature, and in order
to avoid confusion, we omit the labels $(n,l,m)$ in the radial functions.

\begin{figure}[h]
  \centering
    \includegraphics[width=0.5\textwidth]{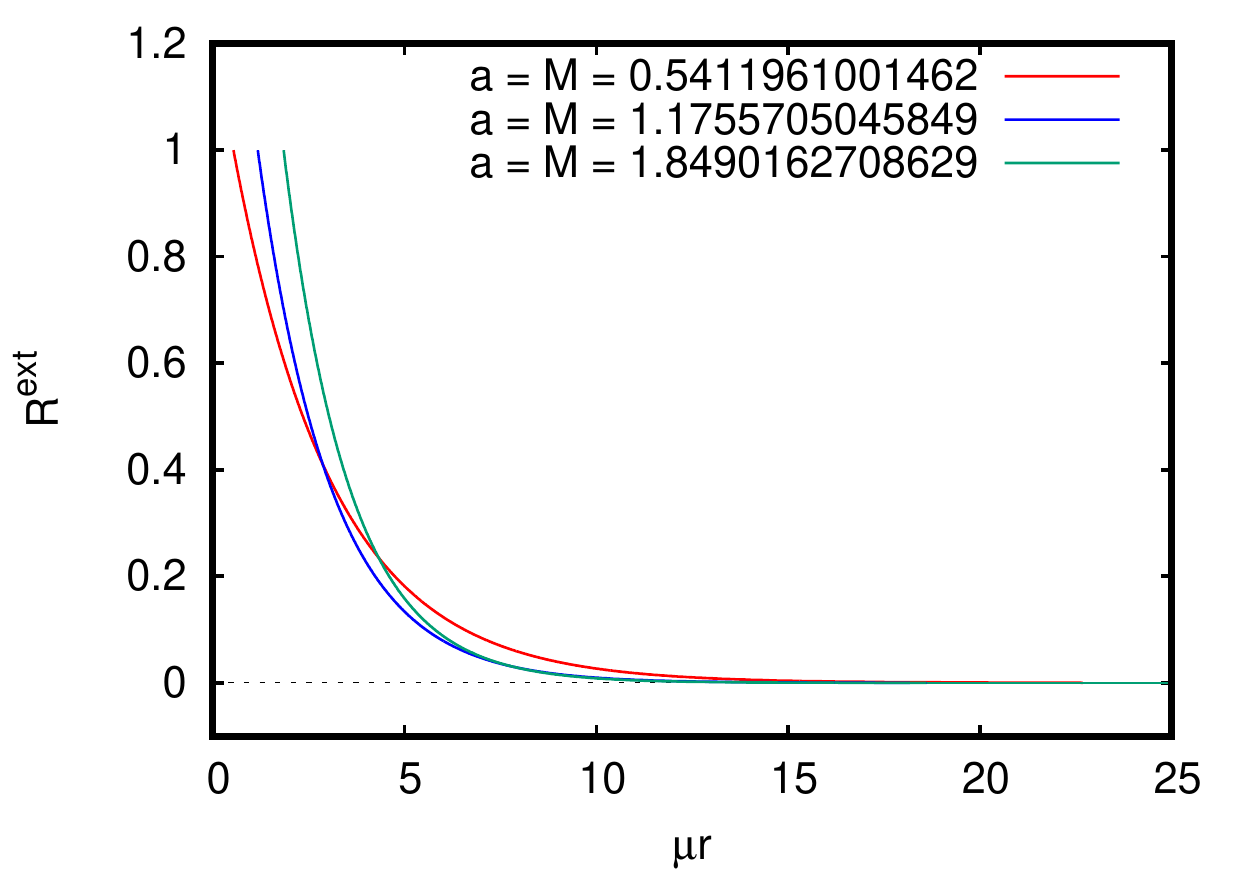}
  \caption{Similar to Fig.~\ref{fig:R11} with no nodes ($n = 0$). The red, blue and green lines correspond to values $m = 1,2,3$, respectively. 
The corresponding eigenvalues $a=M$ are included.} 
  \label{fig:boson}
\end{figure}

Figure~\ref{fig:R11} depicts some examples of radial solutions $R^{\rm ext}$ ($m=1$) 
with different nodes $(n = 0, 1, 2)$ in exact extremal Kerr backgrounds $(a = M)$. 
Figure~\ref{fig:boson} shows a sample of radial solutions without nodes ($n = 0$) for $m=1,2,3$. 

\begin{figure}[h]
  \centering
    \includegraphics[width=0.5\textwidth]{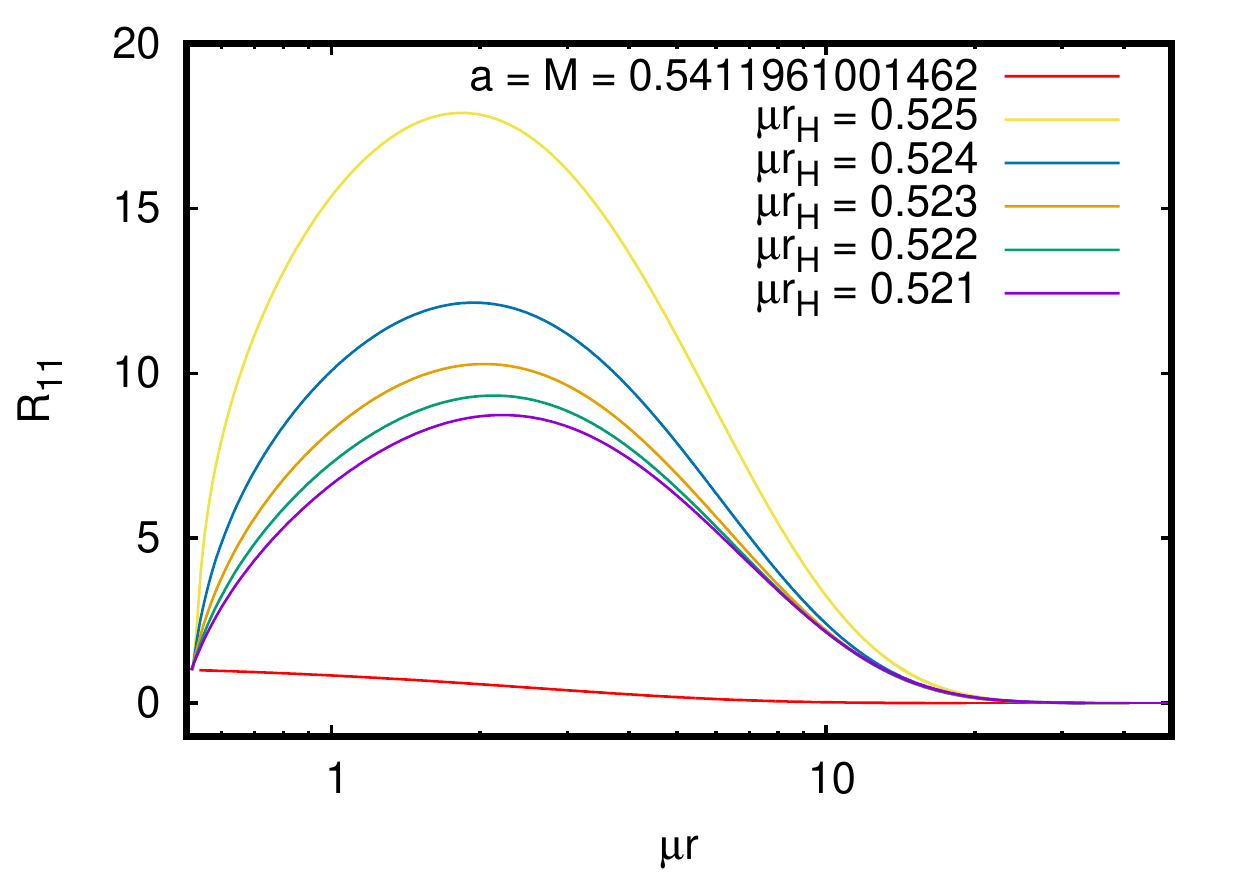}
  \caption{Radial solutions $R$ for the mode $n = 0$ and $m = l = 1$ associated with boson clouds for the near extremal case $r_H \approx M \approx a$ 
    (lines purple, green, orange, blue and yellow) and the exact extremal case $a = M$ (red line; cf. red lines of Figs.~\ref{fig:R11} and~\ref{fig:boson}). 
The location of the horizon $\mu r_H$ is displayed 
in each case. Notice that the exact superregular extremal configuration (red line) is not connected continuously with the near extremal ones whose gradients diverge 
when $r_H \rightarrow M$. The eigenvalues for $a\approx M$ with $n=0$ and $l=m=1$ are included in Table I of \cite{Garcia2019}. }
  \label{fig:radial}
\end{figure}

Figure~\ref{fig:radial} contrasts the radial solutions for $n = 0$ and $m = l = 1$ in the near extremal scenario analyzed in~\cite{Garcia2019}
with the superregular extremal case. We can appreciate that 
for the configurations approaching extremality from the subextremal case
 the maximum amplitude of the radial function increases as well as the slope at the horizon instead of decreasing,
this can be understood by looking at the regularity conditions imposed at the horizon for the subextremal scenario  (see Eqs.(77) and (78) of Ref.~\cite{Garcia2019}) 
where the radial gradients diverge at $r_H= r_H^{\rm ext}=M$. However, by construction, 
the superregular radial configurations for the exact extremal scenario $ r_H^{\rm ext}=a = M$ have bounded radial gradients at the horizon.

\begin{figure}[h]
  \centering
    \includegraphics[width=0.5\textwidth]{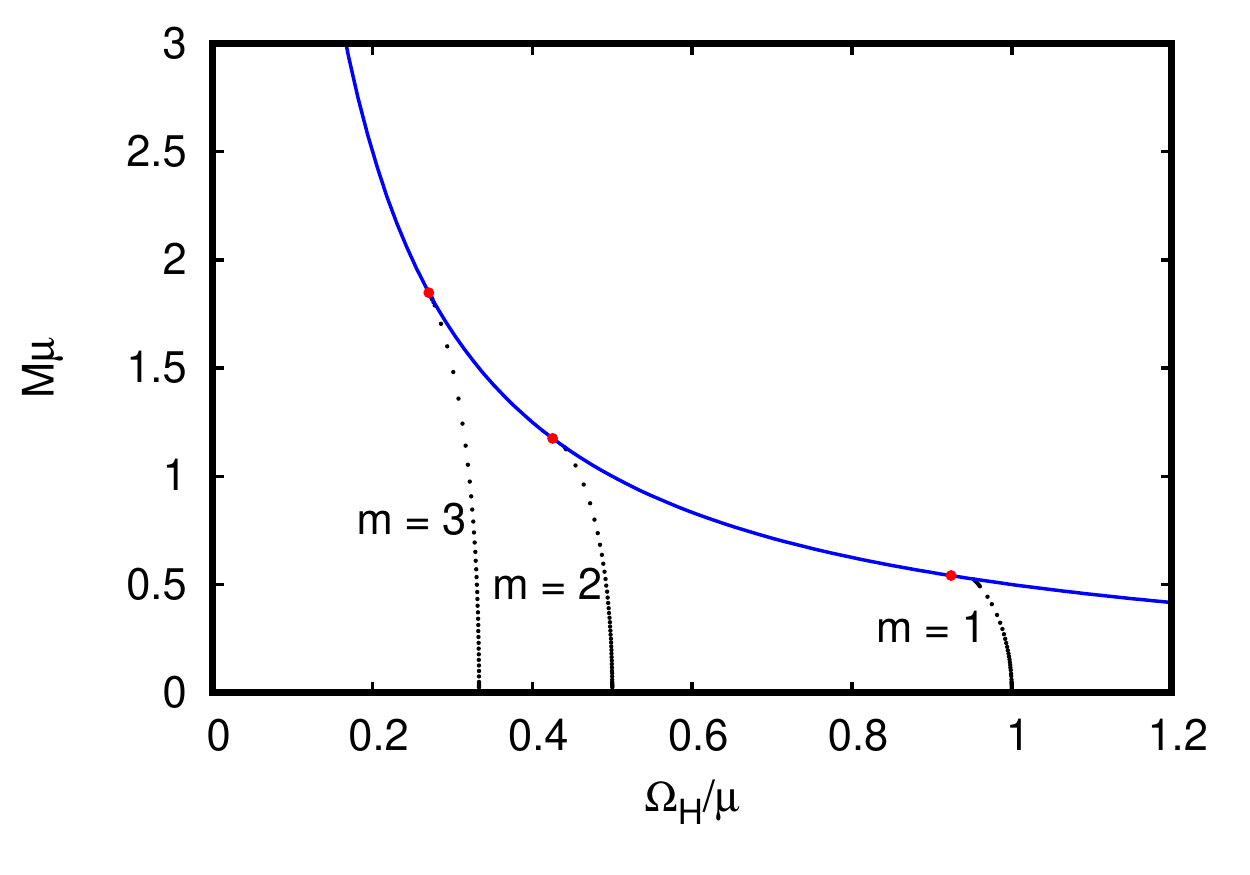}
  \caption{The dotted black lines represent the values for 
the mass $M$ and angular velocity $\Omega_{H}$ 
(in units of $1/\mu$ and $\mu$ respectively) of the Kerr metric that allow for the existence of boson clouds in the {\it subextremal} case. These values 
are found from the eigenvalues $a_{nlm}$ associated with the fundamental mode $n=0$ and 
$l=m$ with $m=1,2,3$ leading to a localized solution for the radial function $R_{nlm}$ (see Ref.\cite{Garcia2019}).
The black dots close to the blue curve represent the specific values of $M, \Omega_H$ for which boson clouds exist in the {\it near extremal} situations. 
The blue solid curve represents the {\it extremal} case $M= 1/(2\Omega_{H})=a$ (Kerr solutions do not exist above this line).
The red dots represent the values obtained for the exact extremal case with $n=0$ and $m=1,2,3$, with radial solutions depicted in Figure~\ref{fig:boson}.} 
\label{fig:MvsOm}
\end{figure}

In Fig.~\ref{fig:MvsOm} we display the {\it existence lines} of boson clouds in a $M$ versus $\Omega_{H}$ diagram for
a background of subextremal and extremal Kerr BH considering $n = 0$ and values $l=m = 1, 2, 3$~\cite{Garcia2019}.
The superregular cloud configurations are associated with the large red dots which, as we notice, are separated from the near extremal dots 
(black dots near the blue line that indicates the extremality relationship Eq.(\ref{OmH})). This feature also indicates that the 
superregular clouds are not continuously connected with the regular extremal ones possessing unbounded radial gradients at the horizon.

\begin{figure}[h]
  \centering
\includegraphics[width=0.5\textwidth]{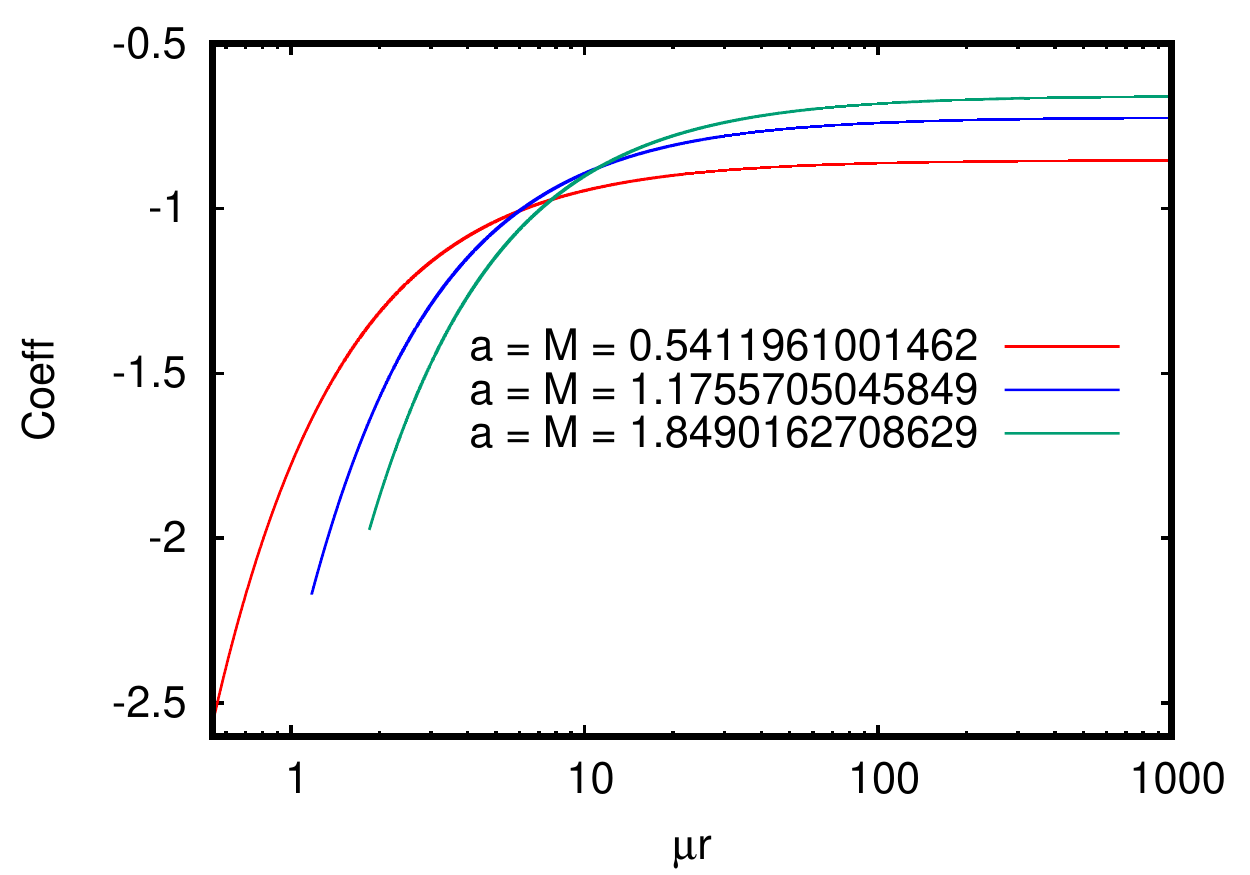}
  \caption{Coefficient $m^2{\cal R}$ associated with the rotational contribution to the 
kinetic term $K=(\nabla_c \Psi^*) (\nabla^c\Psi)$ [see Eqs.~(\ref{kinstataxi}) and (\ref{haircond})]. The coefficient 
is associated with the {\it superregular} radial solutions of Figure~\ref{fig:boson} (i.e. solutions  with $n = 0$)
and evaluated at $\theta=\pi/2$: $m = 1$ (red line), $m = 2$ (blue line), and $m = 3$ (green line). 
The plot shows that $m^2{\cal R}$ is negative in the domain of outer communication: from 
the horizon, $m^2{\cal R}_H = -\frac{3m^2}{4M^2}$, to the asymptotic value, $m^2{\cal R}\rightarrow -\frac{m^2}{4M^2}$
(cf. Table~\ref{tab:kin}).}
  \label{fig:Kinetic}
\end{figure}

Figure~\ref{fig:Kinetic} depicts the coefficient $m^2{\cal R}$ given by Eq.(\ref{haircond}), which is associated with 
the kinetic term Eq. (\ref{kinstataxi}), computed at $\theta = \pi/2$ for simplicity, with the 
parameters associated with the radial solutions shown in Figure~\ref{fig:boson}.
At the horizon ${\cal R}_H= -3/(4M^2)<0$ and asymptotically ${\cal R}\sim -\Omega_H^2=-1/(4M^2)$ \cite{Garcia2019}, which are both negative 
(cf. Table~\ref{tab:kin}). We appreciate that this quantity is negative in the exact extremal case, unlike what happens in the subextremal $(a < M)$ and 
the near extremal scenarios $(a \approx M)$, where the rotational part is positive at and near the horizon and then becomes negative (see Fig.6 in Ref.~\cite{Garcia2019}). 
This behavior indicates that the inequality (\ref{haircond}) actually holds in the domain of outer communication of the BH in the 
exact extremal (superregular) case.

\begin{table}[h]
\centering
\begin{tabular}{|c|c|c|c|}
\hline
 $m$ & $r_H^{\rm ext}$ & $m^2{\cal R}\left(r_{H}^{\rm ext}\right)$ & $m^2{\cal R}\left(r \rightarrow \infty \right)$\\
\hline 
\multirow{1}{0.2cm}{1} & 0.5411961001462 & -2.5606601717797 & -0.8535533906856\\ \hline
\multirow{1}{0.2cm}{2} & 1.1755705045849 & -2.1708203932501 & -0.7236067979201\\ \hline
\multirow{1}{0.2cm}{3} & 1.8490162708629 & -1.9743416490252 & -0.6581138832518\\ \hline
\end{tabular}
\caption{Values of the coefficient $m^2{\cal R}$ at $\theta=\pi/2$ computed at the horizon $r_H^{\rm ext}$ (in units of $1/\mu$) 
and asymptotically. 
These values are associated with Figure~\ref{fig:Kinetic}.}
\label{tab:kin}
\end{table}

Something remarkable that we did not realize initially during our numerical computations, but which incidentally corroborates the accuracy of 
the numerical analysis, is that the spectra for $a$ can be obtained exactly and in closed form from Hod's resonance condition (\ref{kappasmooth}) and 
from Eq.~(\ref{kappaHod}):
\begin{equation}
  \label{spectra}
M=a= \frac{1}{2\mu}\sqrt{m^2 + \left[-\kappa+\sqrt{\kappa^2+m^2}\,\right]^2}\;.
  \end{equation}
This is a notable formula. Notice that it does not depend explicitly on the number $l$. This dependence is 
implicit in $\kappa$ via the quantity $\beta$ in Eqs. (\ref{kappasmooth0}) and (\ref{betaHod}).

 We checked that the spectra found numerically using a shooting method
match perfectly well the spectra found from Eq.~(\ref{spectra}), with $\kappa$ provided by (\ref{kappasmooth}). Tables \ref{tab:MOmega} and \ref{tab:MOmega2} include some examples of this concordance.
 
\begin{table}[htbp]
\begin{center}
\begin{tabular}{|c|c|c|c|}
\hline
$\mu r_{H}$ & $\mu M_{\rm num}$ & $\mu M_{\rm ana}$\\
\hline \hline
0.5411961001462 & 0.5411961001462 & 0.5411961001461\\ \hline
0.5137430040861 & 0.5137430040861 & 0.5137431483730\\ \hline                    
0.5065407258909 & 0.5065407258909 & 0.5065407286165\\ \hline
\end{tabular}
\caption{Numerical (using a shooting method) and analytic [using Eq.(\ref{spectra})] values of $\mu M$ for $m = 1$ and principal number $n = 0, 1, 2$ associated with Figure~\ref{fig:R11}.}
\label{tab:MOmega}
\end{center}
\end{table}

\begin{table}[htbp]
\begin{center}
\begin{tabular}{|c|c|c|c|}
\hline 
$\mu r_{H}$ & $\mu M_{\rm num}$ & $\mu M_{\rm ana}$\\
\hline \hline 
0.5411961001462 & 0.5411961001462 & 0.5411961001461\\ \hline
1.1755705045849 & 1.1755705045849 & 1.1755705045849\\ \hline                     
1.8490162708629 & 1.8490162708629 & 1.8490162708629\\ \hline
\end{tabular}
\caption{Numerical (using a shooting method) and analytic [using Eq.(\ref{spectra})] values of $\mu M$ for the fundamental mode $(n = 0)$ with $m = 1, 2, 3$ associated with Figure~\ref{fig:boson}. }
\label{tab:MOmega2}
\end{center}
\end{table}

\begin{figure}[h!]
\begin{center}
\includegraphics[width=0.45\textwidth]{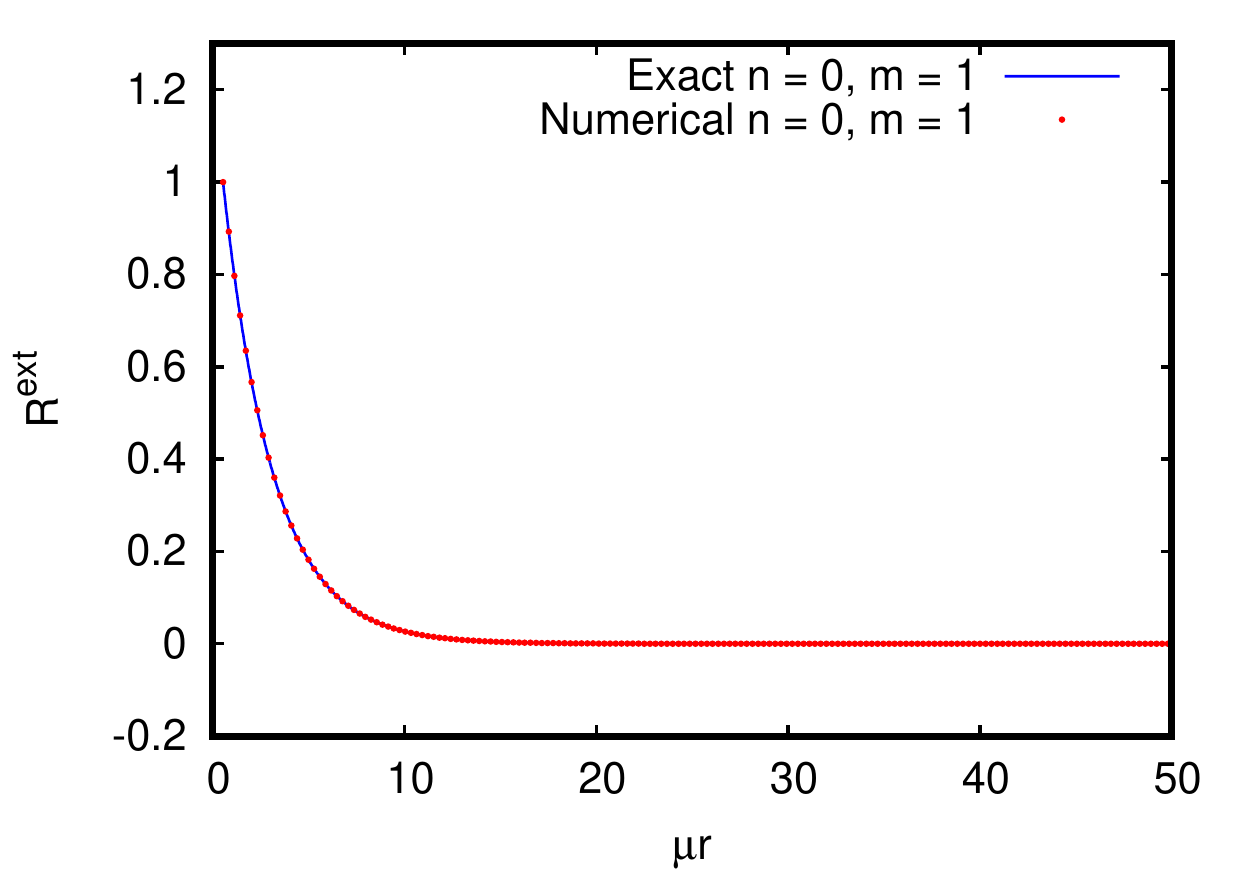}
\includegraphics[width=0.45\textwidth]{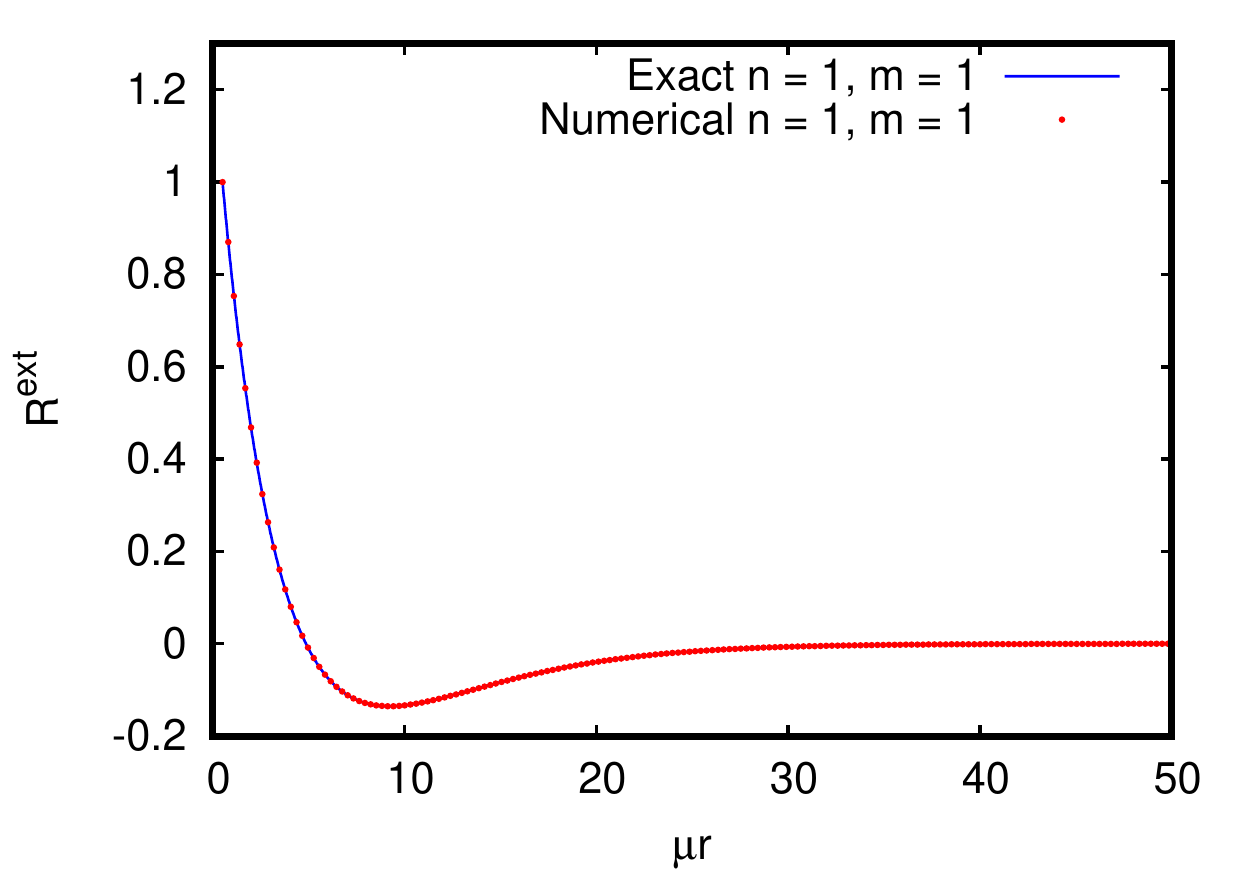}
\includegraphics[width=0.45\textwidth]{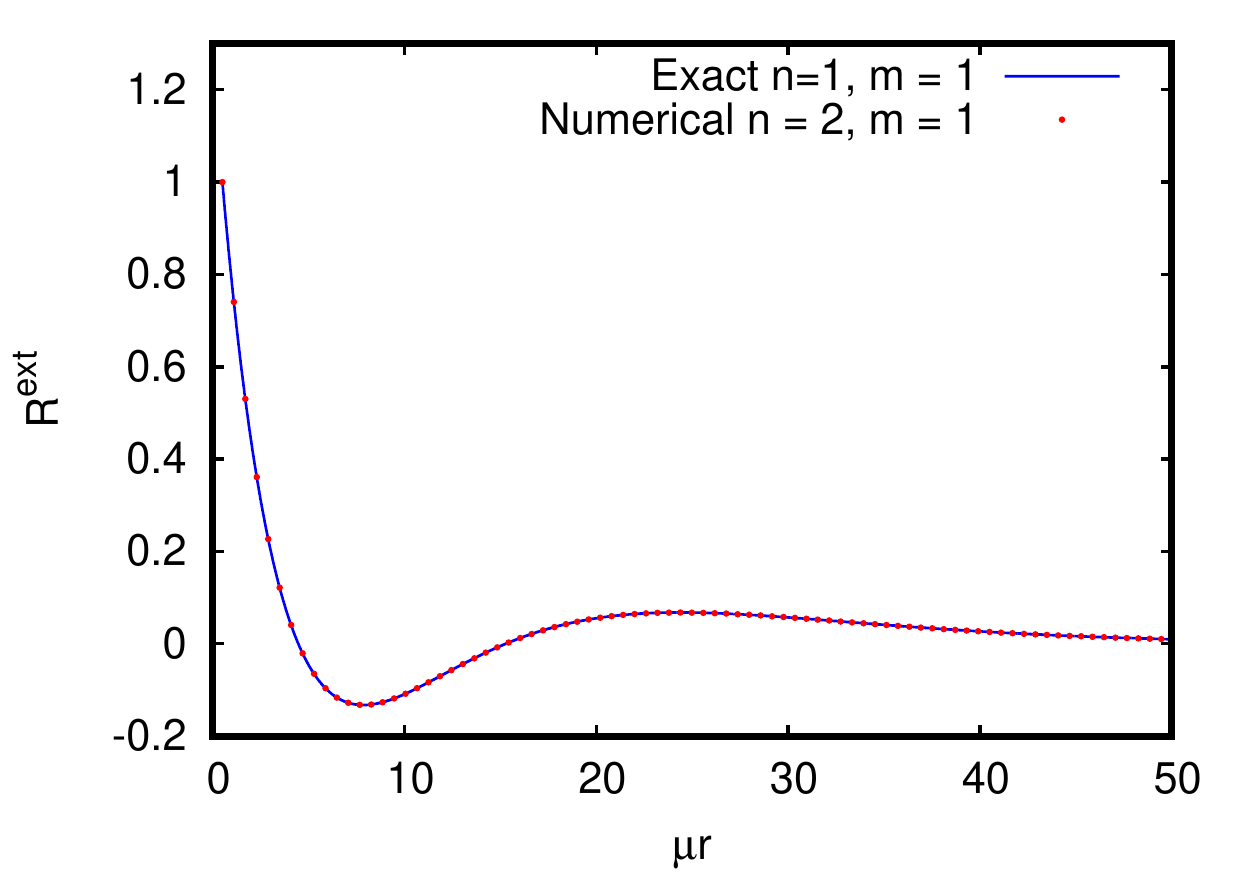}
\caption{Radial solutions $R^{\rm ext}$ associated with the extremal clouds with principal numbers $n = 0, 1, 2$ and $m = 1$. 
The red dots correspond to the numerical solutions we obtained by solving Eq. (\ref{radialE}) and the blue continuous line corresponds to the exact 
solution given by Hod (\ref{RHod2}) with $\beta=1/2$, and the constant $A$ is chosen so that $R_H=1$ for each $n$. 
Notice the perfect agreement between both type of solutions.}\label{fig:HodNumerical}
\end{center}
\end{figure}

Using Eq.~(\ref{spectra}) we then corroborated that for other values for $n,m$ one obtains 
readily the localized radial functions numerically without the use of any shooting method at all. This is perhaps one of the few cases, if not the only
one, where the spectra of an eigenvalue problem is found analytically and in closed form when taking into account gravity. 
Figure~\ref{fig:HodNumerical} compares our numerical solution for the radial function using $r_H=M=a$ from Eq.~(\ref{spectra})
with that obtained by Hod analytically (\ref{RHod2})~\cite{Hod2012} taking $\beta=1/2$ for $m = 1$ and different values of $n$.

\begin{figure}[h]
  \centering
    \includegraphics[width=0.5\textwidth]{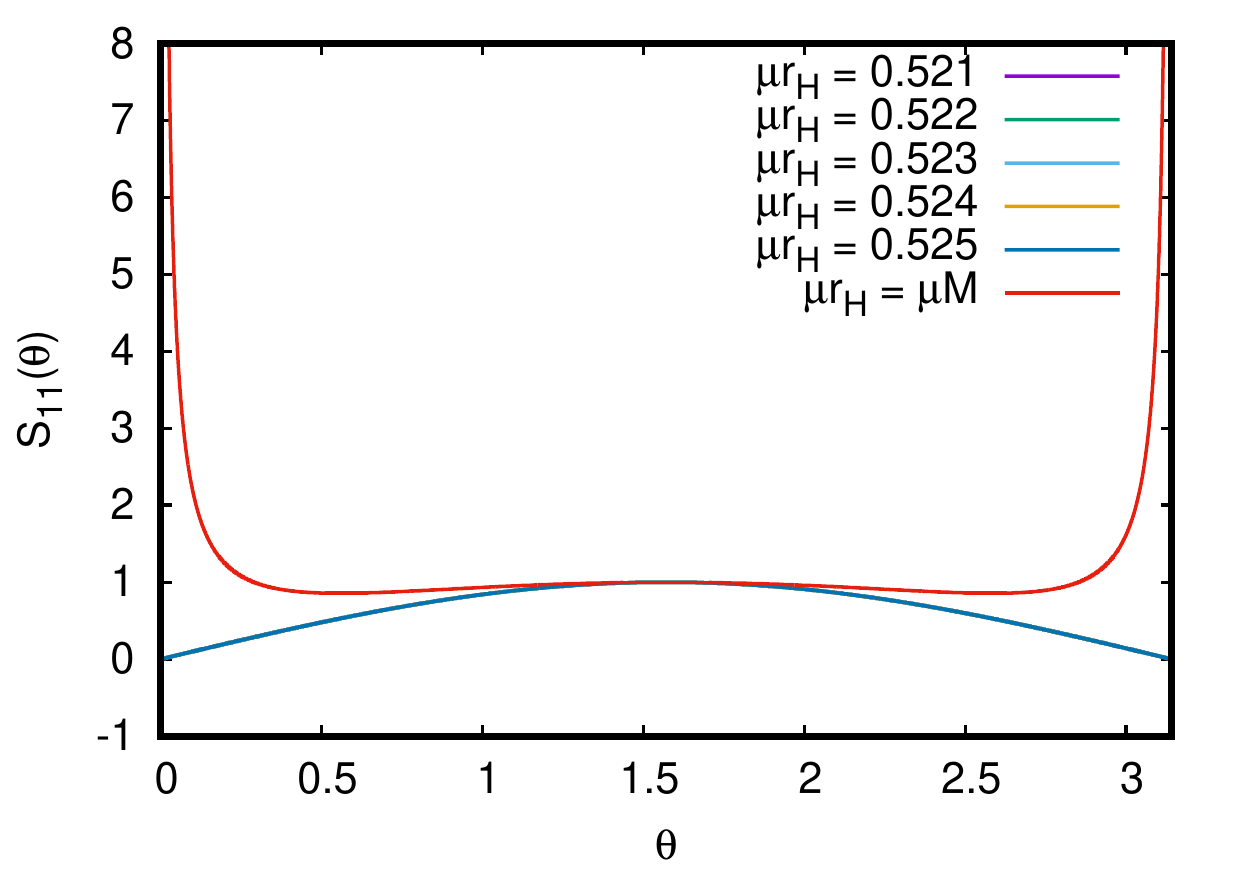}
  \caption{Spheroidal harmonic (SH) $S_{lm}(\theta)$ for $n = 0$ and $m =l= 1$ for the near extremal $r_H\approx a \approx M$ 
(colored lines superposed) and extremal $r_H=a = M$ (red line) cases, respectively. For the near extremal scenario 
(the horizon location is indicated), we use the separation constants given by (\ref{Klmsubext}), 
and for the exact extremal case $(a = M)$ associated with $\beta=1/2$ we use $K_{m}^{\rm ext}$ given by (\ref{Klm}). 
Notice the divergent behavior of the SH at $\theta=0,\pi$ in the exact extremal case (red line) and their adequate behavior in the near extremal
scenario (colored lines other than the red one, all superposed). The divergent behavior in the exact extremal scenario is because
the separation constants $K_{m}^{\rm ext}$ do not allow regularity on the axis of symmetry. The resulting 
SH is not an acceptable solution 
despite the fact that the corresponding radial function is perfectly well behaved (cf. upper panel of Figure~\ref{fig:HodNumerical}).}
\label{fig:Angular}
\end{figure}

\begin{figure}[h!]
\begin{center}
\includegraphics[width=0.35\textwidth]{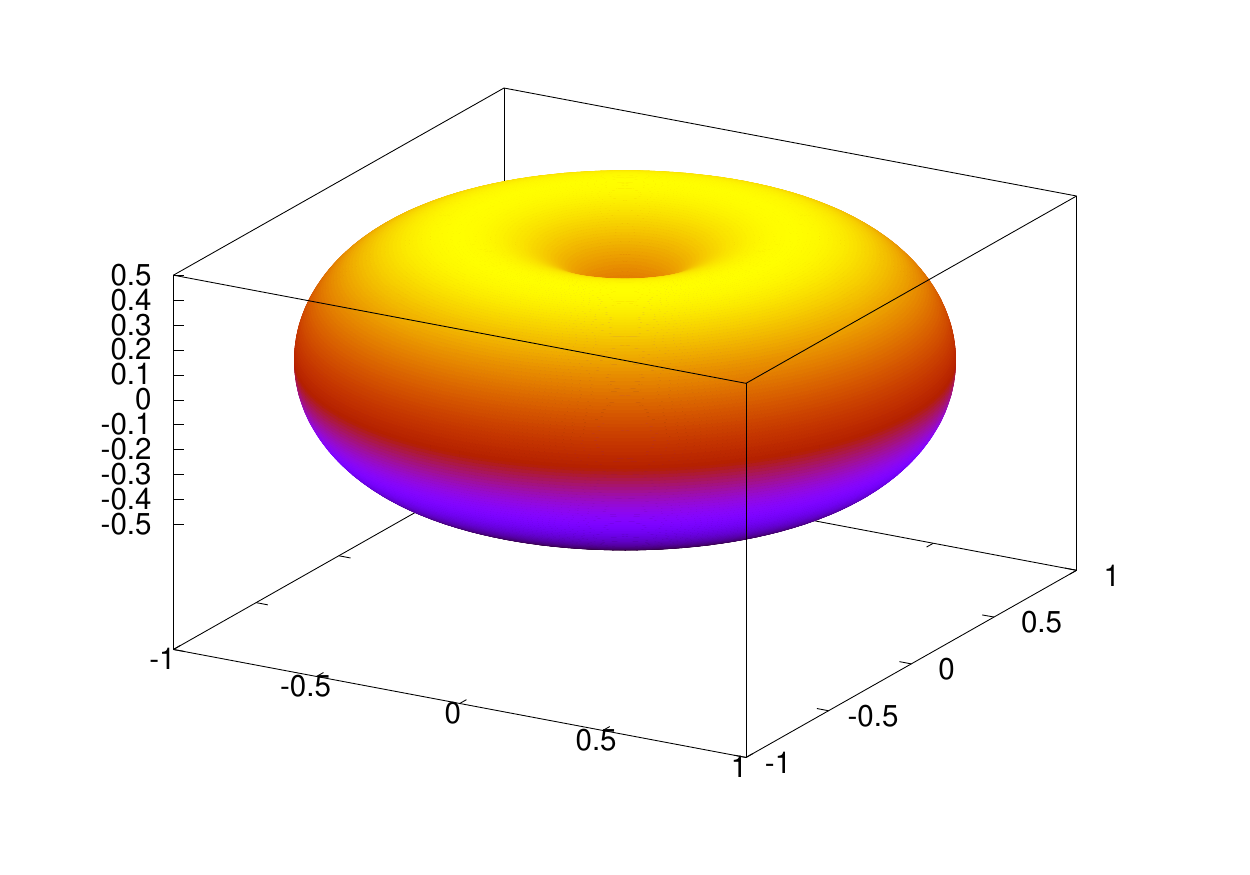}
\includegraphics[width=0.35\textwidth]{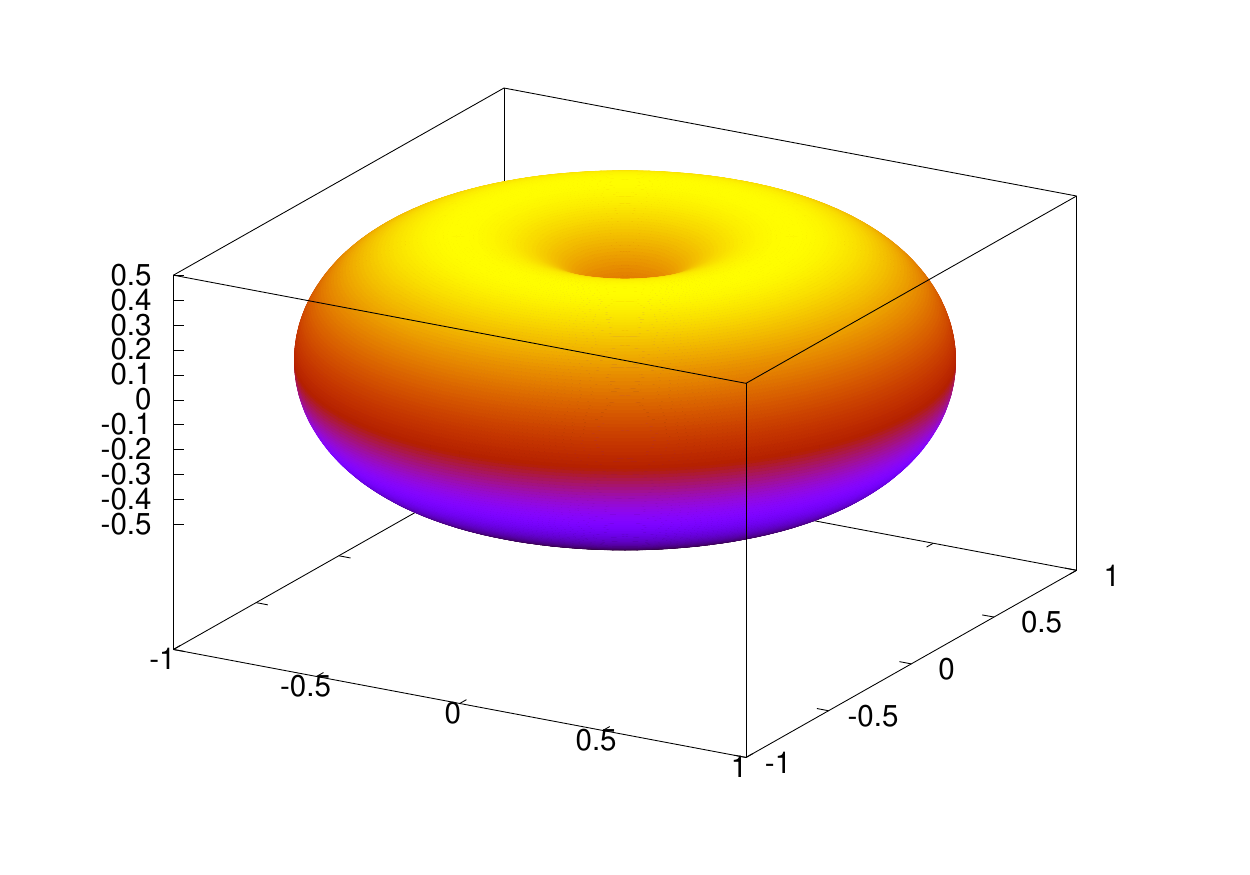}
\includegraphics[width=0.35\textwidth]{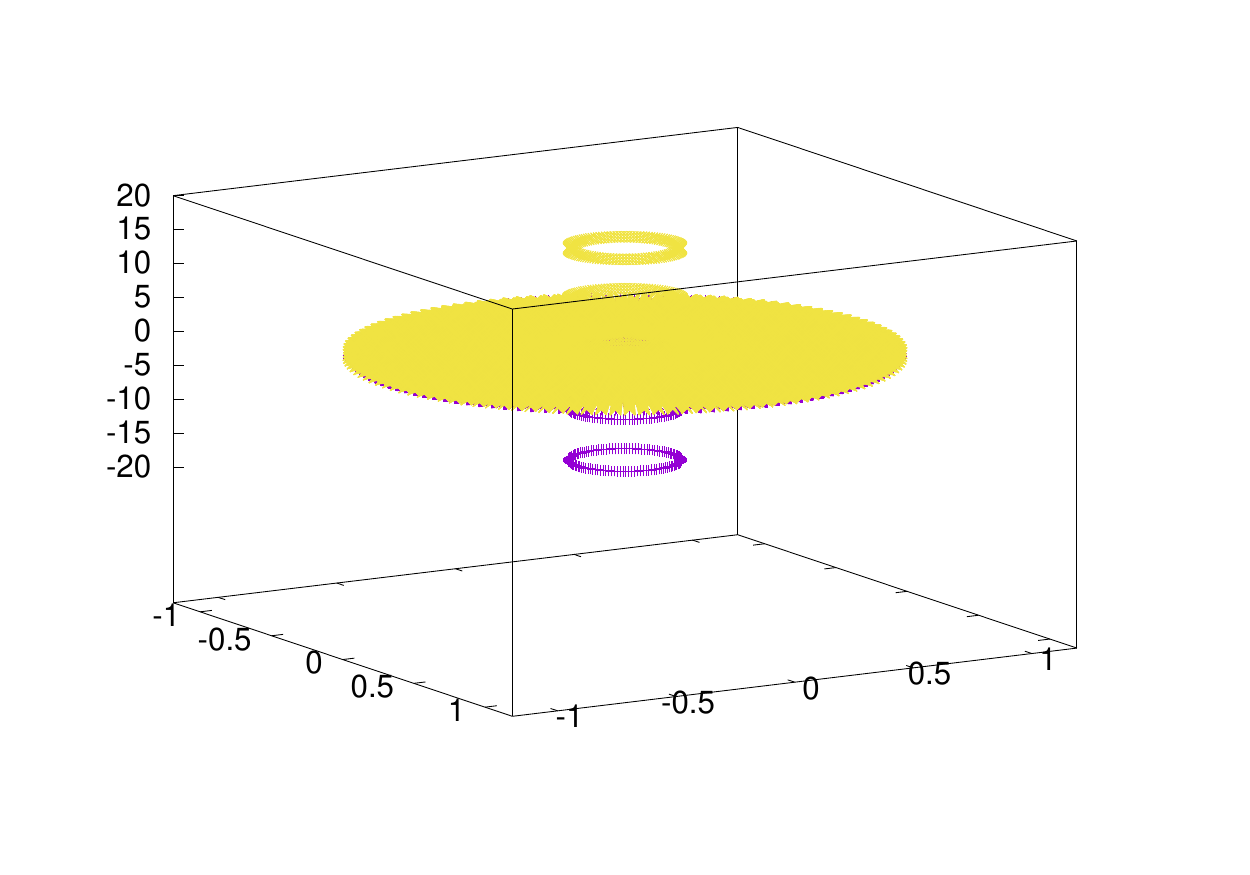}
\caption{Spheroidal harmonics $|S_{lm}(\theta,\varphi)|$ associated with Figure~\ref{fig:Angular} ($n = 0$ and $m =l= 1$). 
The top panel and the middle panel correspond to the near-extremal  case with $\mu r_{H} = 0.523$ and $\mu r_{H} = 0.525$, respectively. 
The bottom panel corresponds to the exact extremal case with $\mu r_{H} = 0.5411961001462 = \mu M= \mu a$. 
 For the near extremal and exact extremal scenarios we use the separation constants (\ref{Klmsubext}), and (\ref{Klm}), 
respectively. Like in Figure~\ref{fig:Angular} we appreciate the pathological behavior of $|S_{lm}(\theta,\varphi)|$ at $\theta=0,\pi$ for the 
exact extremal case. }\label{fig:Sm}
\end{center}
\end{figure}

Unfortunately these ``astonishing'' findings are not fully consistent. If in addition one tries to solve numerically
the eigenvalue problem for the SH (\ref{angularE}) using the separation constants (\ref{Klm}) and the eigenvalue $a=M$ found
numerically or analytically using Eq.~(\ref{spectra}), the resulting $S_{lm}$ turns out to be very badly behaved 
(i.e. the SH is divergent) at $\theta=0,\pi$ (i.e. on the axis of symmetry) as it is shown
in Figure~\ref{fig:Angular} (red line) and Figure~\ref{fig:Sm} (lower panel) for $m = 1$.  
This is in contrasts with the acceptable behavior on the axis of symmetry of a similar SH for the near extremal case 
where the separation constants $K_{lm}$ are given only by (\ref{Klmsubext}). 
A sample of well behaved SH in the near extremal case are depicted in Figure~\ref{fig:Angular} (colored lines 
other than the red one; those colored lines are superposed) and in the upper and middle panels of Figure~\ref{fig:Sm}.
Conversely, if one employs the separation constants (\ref{Klmsubext}) that provides well behaved SH at $\theta=0,\pi$ to find the solution for the 
radial function, the latter becomes bad behaved at the horizon.

In the specific exact extremal scenario with $\beta=1/2$ the separation constants (\ref{Klm}) may correspond to 
$l=0$, notably when $n\rightarrow \infty$ (see the discussion below)\footnote{One can write the hydrogen-atom radial wave function that is regular at $r=0$ in terms of a confluent hypergeometric function with a $\beta= l + 1/2$ having exactly the same form as 
Hod's exact solution (\ref{RHod}). Thus, the value $\beta=1/2$ corresponds to $l=0$, which implies $m=0$. In this case the solution is 
a spherically symmetric wave function. We acknowledge C. Herdeiro for this remark. Furthermore, such solution and its first and second derivatives are also non-zero at $r=0$. Those wave functions are allowed to exist in the case of the hydrogen atom, however, as stressed in the main text, in the current scenario the 
value $m=0$ leads to vanishing clouds as they cannot exist in spherical symmetry under the assumptions adopted here.}.
 If this is the case consistency requires to take $m=0$ since $|m|\leq l$, leading then to a contradiction with the 
fact that the corresponding non-trivial radial functions only exist for $m\neq 0$. 
On the other hand, if one takes $m=0$ (a purely spherically symmetric mode)
then the clouds disappear as in that case the only solution for the field is $\Psi\equiv 0$, a conclusion that is consistent with
the no-hair theorems in asymptotically flat spacetimes that states that in spherically symmetric situations the boson field must vanish.

One can then wonder if it is possible to find non-trivial values for $m$ and $l$ such that one can reconcile both types of separation constants
(i.e. $K_m^{\rm ext}= K_{lm}$), and thus to obtain acceptable SH at at $\theta=0,\pi$ together with superregular radial functions at the horizon. 
The best we could do in this direction was to consider a scenario where $n\gg m \neq 0$. According to 
Eq.~(\ref{kappasmooth}) this assumption implies $\kappa \approx n$. We can then expand $M$ in Eq.(\ref{kappasmooth}) for $\kappa\gg m$,
\begin{eqnarray}
  \label{spectra1}
  M\mu = \mu a &=&  \frac{m}{2}+ \frac{m^3}{16\kappa^2} + {\cal O}\left(\frac{m^5}{\kappa^4}\right) \nonumber \\
  &\approx& \frac{m}{2}+ \frac{m^3}{16 n^2} +  {\cal O}\left(\frac{m^5}{n^4}\right) \;,
\end{eqnarray}
which was found by Hod~\cite{Hod2012} for the regular extremal clouds assuming $\beta\geq 1/2$ and for $n\gg l$~\footnote{Actually, Hod's Eq.~(22) reads
$M\mu_{\pm}= \frac{m}{2}+ \frac{m^3}{16 n^2} - \frac{m^3}{16 n^3}(1\pm 1) + {\cal O}(n^{-4})$. Thus the subleading terms beyond the second 
one do not coincide with Eq.~(\ref{spectra1}). For $\beta=1/2$ we find one kind of resonance which presumably is associated with $M\mu_{+}$.}.
Thus, if we take the leading term, which is the only one that survives in the limit $n \rightarrow \infty$, 
we obtain $M\mu\approx m/2$. Equations (\ref{OmH}) and (\ref{fluxcond2}) yield $\omega= m\Omega^{\rm ext}_H= \frac{m}{2M}$, 
and in this limit $\omega \approx \mu$. It is interesting to note that in this limit the value $M\mu\approx m/2$ 
corresponds to the lower bound for $M$ given by the band (\ref{Mband}). 

Using these results, we find that the separation constants (\ref{Klmsubext}) 
become independent of $m$ and $M$, and reduce simply to
\begin{equation}
\label{Klmexp}
K_{lm} = l(l + 1)\;,
\end{equation} 
while the separation constants Eq.(\ref{Klm}) read,
\begin{equation}
\label{Klm1}
K_{m,\frac{1}{2}}^{\rm ext} = \frac{3}{2}m^2\;.
\end{equation} 
Thus, compatibility for both the angular and radial parts of the boson field requires that the separation constants 
(\ref{Klmexp}) and (\ref{Klm1}) coincide. This condition leads to the following relationship\footnote{In Ref.[43] of \cite{Hod2012}, Hod 
reports an inequality in the limit $n\rightarrow \infty$. When the inequality is saturated it coincides with condition (\ref{mlninf}).}

\begin{equation}
\label{mlninf}
m = \sqrt{\frac{2l(l + 1)}{3}}\;.
\end{equation} 

Equation (\ref{mlninf}) can be rewritten as 
\begin{equation}
(2l + 1)^2 - 6m^2 = 1 \;,
\end{equation}
which takes the form of a quadratic Diophantine equation, also known as Pell's equation:
\begin{equation}
\label{Pell1}
x^2 - Dy^2 = 1 \;,
\end{equation}
with 
\begin{eqnarray}
&& D = 6 \;,\\
\label{pairsxy}
&& x = 2l + 1 \;,\; y= m \;.
\end{eqnarray}
Equation (\ref{Pell1}) admits the trivial solution $x=1,y=0$ (i.e. $l=0=m$). According to 
number theory, when $D$ is a positive nonsquare integer 
(see \cite{Barbeau2003,Whitford1912,Lenstra2002} for a review), it is possible to find a sequence of solutions $(x_i, y_i)$ generated from 
the smallest non-trivial solution $(x_0, y_0)$, called the fundamental solution, through the recurrence relations
\begin{eqnarray}
\label{Pellrecx}
x_i &=& x_{i-1}x_0 + y_{i-1}y_{0}D\;,\\
\label{Pellrecy}
y_i &=& x_{i-1}y_0 + y_{i-1}x_0\;.
\end{eqnarray}
By direct substitution of Eqs.~(\ref{Pellrecx}) and (\ref{Pellrecy}) in Eq.~(\ref{Pell1}) one verifies that $(x_i,y_i)$ is a solution
  if $(x_{i-1},y_{i-1})$ is also a solution. In this case the fundamental solution $(x_0, y_0)$ is $(5, 2)$, which corresponds to the values $l = 2$, $m = 2$. 
The fundamental pair $(5, 2)$ generates $(x_1,y_1)=(49, 20)$, which is 
associated with the numbers $l = 24$, $m = 20$. In this way it is possible to generate all the values $l$ and $m$ 
satisfying Eq.~(\ref{mlninf}). Table~\ref{tab:Klmninf} displays a sample of pairs $(x,y)$ and $(l,m)$ 
generated by the fundamental one.

 \begin{table}[htbp]
\begin{center}
\begin{tabular}{|c|c|c|c|}
\hline 
$(x,y)$ & $l$ & $m$ & $K_{m,\frac{1}{2}}^{\rm ext} = K_{lm} $\\
\hline \hline 
(1,0)    & 0 & 0 & 0\\ \hline
(5,2)    & 2 & 2 & 6  \\ \hline                     
(49,20)    & 24 &  20  & 600  \\ \hline
(485,198)   & 242 & 198 & 58806  \\ \hline
(4801,1960) & 2400 & 1960 & 5762400  \\ \hline
(47525,19402)& 23762 & 19402 & 564656406 \\ \hline
\end{tabular}
\caption{Sample of pairs $(x,y)$ (column 1) satisfying Pell's Eq.~(\ref{Pell1}), including the trivial pair $(1,0)$. 
The non-trivial pairs are constructed from the fundamental one $(5,2)$ using  
Eqs.~(\ref{Pellrecx}) and (\ref{Pellrecy}). Columns 2 and 3 tabulate the 
integer values $m$ and $l$ obtained from (\ref{pairsxy}) satisfying the compatibility condition (\ref{mlninf}) associated with the limit 
$n\rightarrow \infty$. The pair $l=0=m$ leads to trivial cloud solutions. Column 4 shows the values of the separation constants
(\ref{Klmexp}) and (\ref{Klm1}).}
\label{tab:Klmninf}
\end{center}
\end{table}

Because numerically is impossible to achieve the limit $n \rightarrow \infty $, Table~\ref{tab:ninfinity} 
displays a specific example for the values of the separation constants as we increase $n$ given $m=2=l$. 
These values are computed as follows: in the case 
of $K^{\rm ext}_{m,\frac{1}{2}}$ we use (\ref{Klm}), and $M$ is obtained from (\ref{spectra}) given the numbers $m,n$. 
As concerns $K_{lm}$, we use (\ref{Klmsubext}) given $n,l,m$ and $M$ from (\ref{spectra}).
 \begin{table}[htbp]
\begin{center}
\begin{tabular}{|c|c|c|}
\hline 
$n$ &  $K^{\rm ext}_{m,\frac{1}{2}}$  & $K_{lm}$ \\
\hline \hline 
10 & 5.9837387624884322  &  5.9930307696716758 \\ \hline
$10^2$ &  5.9998039792199140 & 5.9999159910747437 \\ \hline                     
$10^3$ & 5.9999980039979928 & 5.9999991445705581 \\ \hline
$10^4$ & 5.9999999800039996 & 5.9999999914302853  \\ \hline
$10^5$ & 5.9999999998000044 & 5.9999999999142872\\ \hline
$10^6$ & 5.9999999999979998 & 5.9999999999991429\\ \hline
$10^7$ & 5.9999999999999805 & 5.9999999999999920  \\ \hline
\end{tabular}
\caption{Behavior of separation constants (\ref{Klm}) and (\ref{Klmsubext}) for $m = l = 2$ as $n \rightarrow \infty$. 
These values are compatible with the exact value given in Table~\ref{tab:Klmninf} for $m = l = 2$. }
\label{tab:ninfinity}
\end{center}
\end{table}

\begin{figure}[h]
  \centering
    \includegraphics[width=0.5\textwidth]{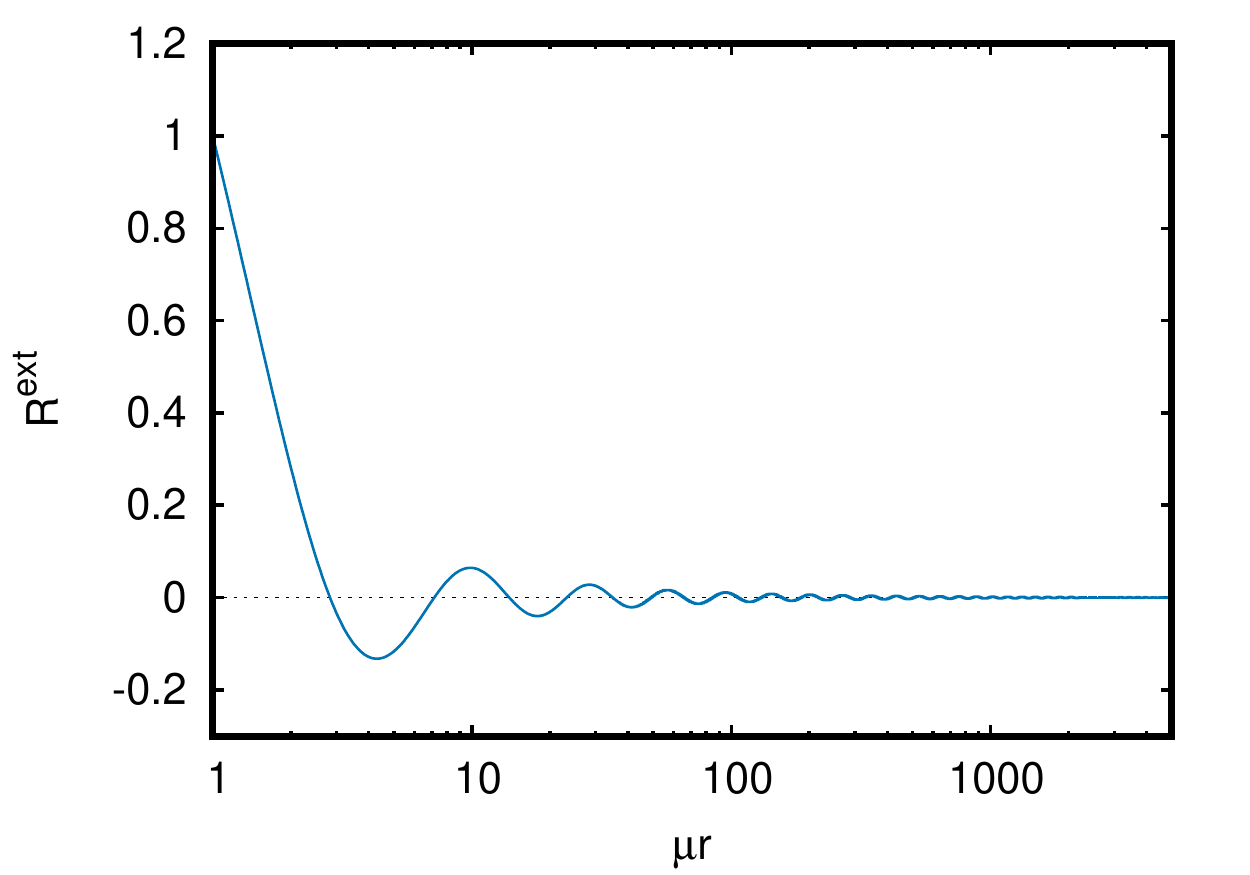}
  \caption{Radial solution $R^{\rm ext}$ with principal number $n = 10^9$ and $m = 2$ with 
$r_H^{\rm ext}=M=a\approx 1/\mu$ [cf. Eq.~(\ref{spectra1})].} \label{fig:RInf}
\end{figure}

\begin{figure}[h!]
\begin{center}
\includegraphics[width=0.45\textwidth]{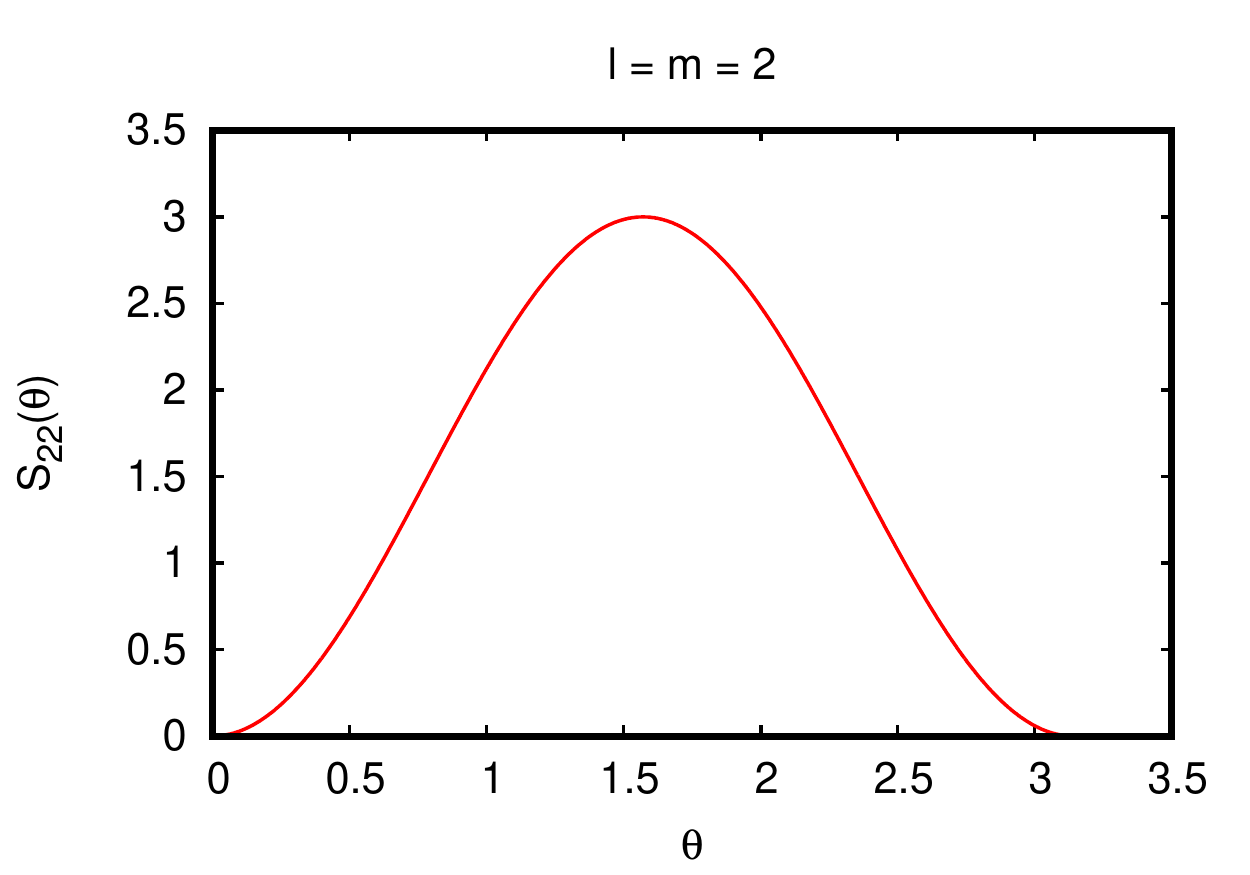}
\includegraphics[width=0.45\textwidth]{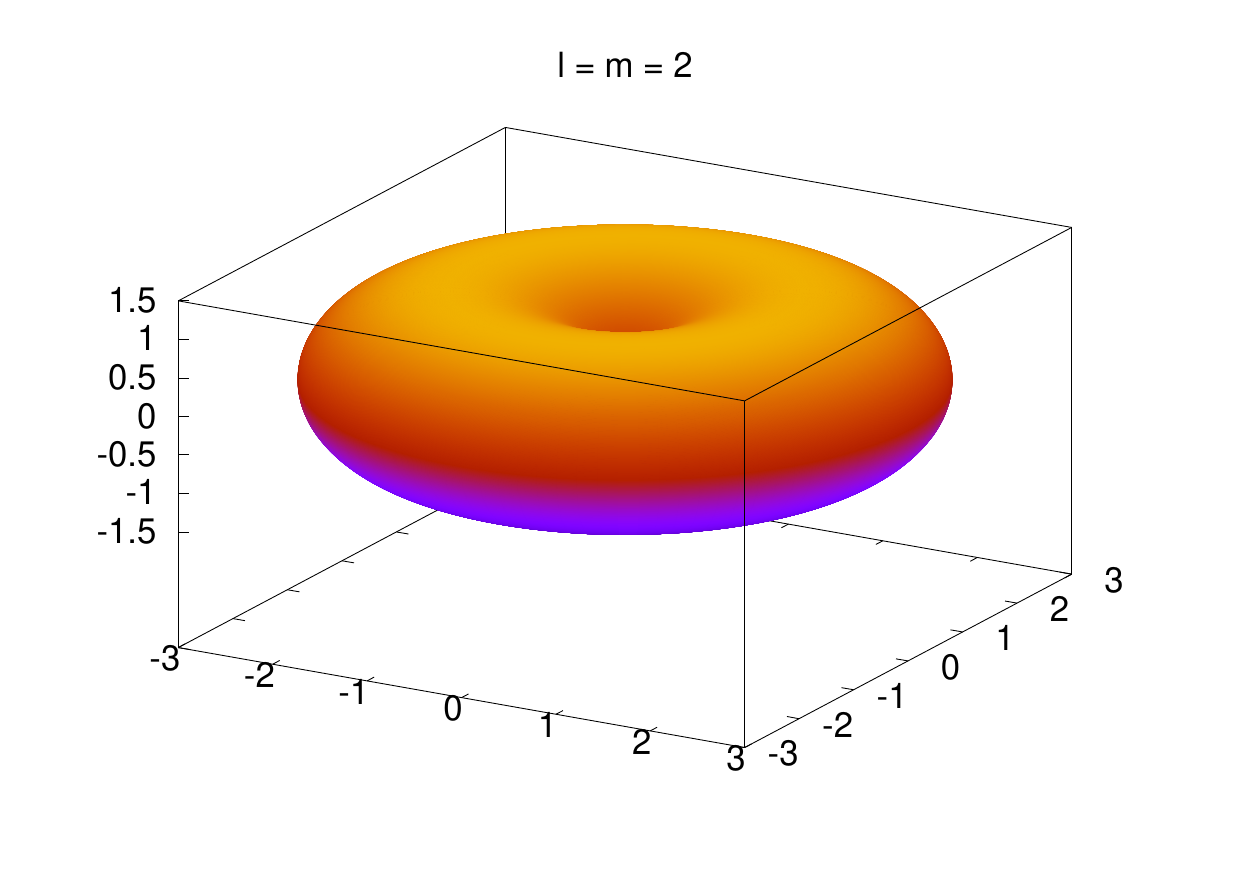}
\caption{Spheroidal harmonic (SH) $S_{lm}(\theta)$ (upper panel) and  $|S_{22}(\theta,\varphi)|$ (lower panel)
for $m = l = 2$ and $n = 10^9$ in the exact extremal case. In this limit ($n\gg 1$) it is possible to find acceptable angular solutions. 
Unlike the SH depicted in Figure~\ref{fig:Sm} for the extremal case ($n=0$), this SH is well behaved at $\theta=0,\pi$. }\label{fig:Sm22}
\end{center}
\end{figure}

\begin{figure}[h!]
\begin{center}
\includegraphics[width=0.45\textwidth]{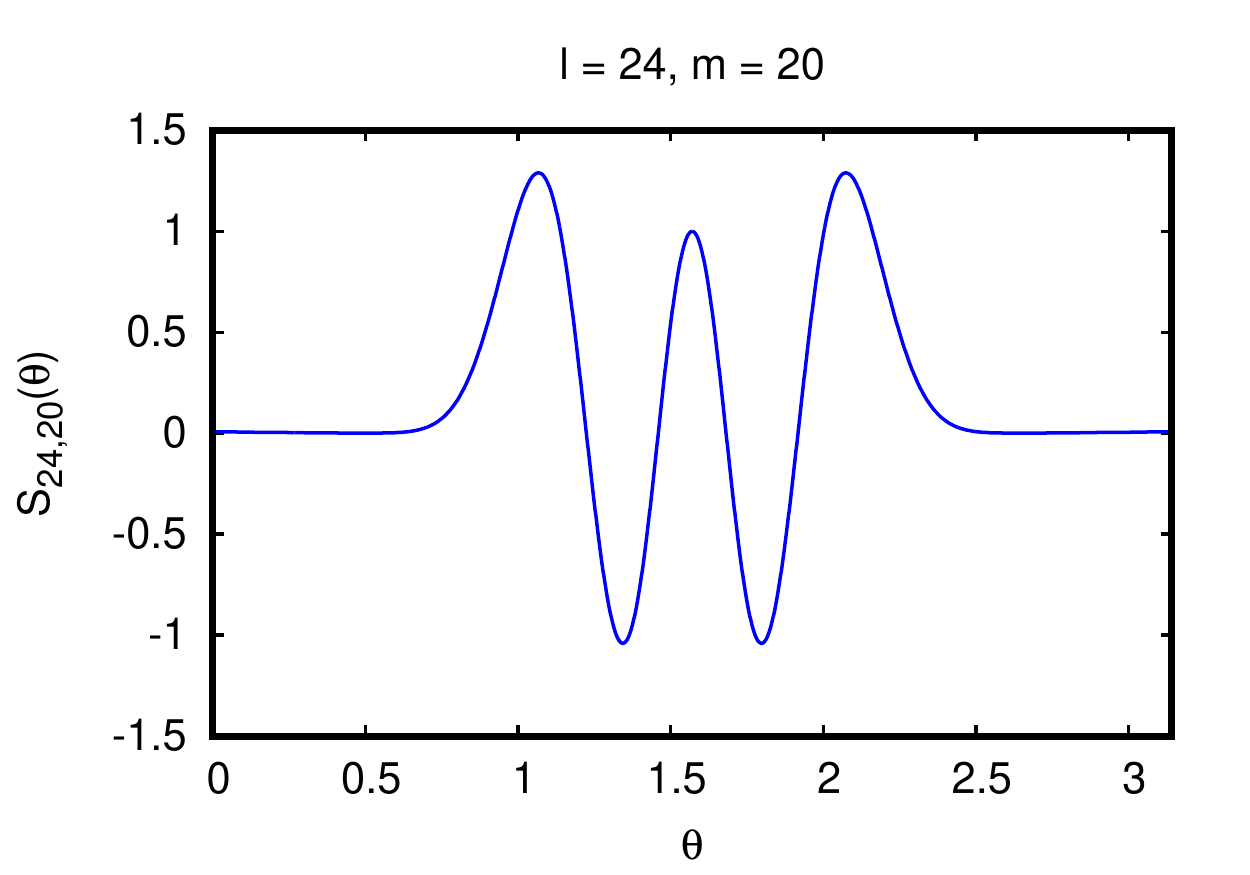}
\includegraphics[width=0.45\textwidth]{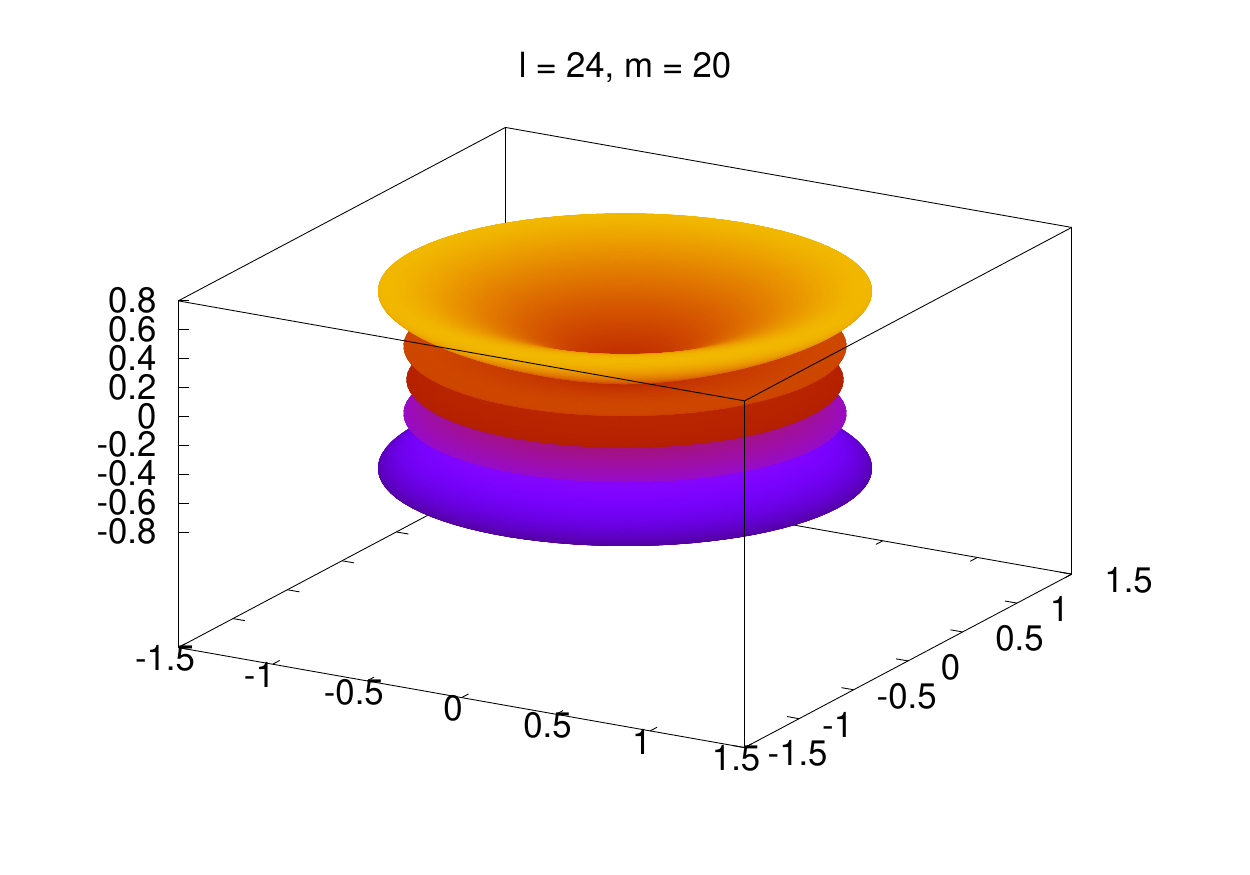}
\caption{Similar to Figure~\ref{fig:Sm22}  for $l = 24$ and $ m = 20$ with $n =10^9$.}\label{fig:Sm2420}
\end{center}
\end{figure}

Figure~\ref{fig:RInf} depicts a radial solution $R^{\rm ext}$ for $m = 2$ and $n = 10^9$, with $M$ provided by Eq. (\ref{spectra}). 
Figures~\ref{fig:Sm22} and \ref{fig:Sm2420} show the SH, $S_{lm}(\theta)$ (upper panels) and  $|S_{lm}(\theta,\varphi)|$ (lower panels) for 
$m = 2$, $m = 20$ and $n = 10^9$, using $K_m^{\rm ext}$ given by Eq.~(\ref{Klm1}), which coincide with $K_{lm}$ (\ref{Klmexp}) when taking $l=2$ and $l=24$, respectively, (see Table~\ref{tab:Klmninf}), 
and $\mu r_H=\mu M=\mu a=m/2$. Since in these plots $n\gg 1$ it is possible to find acceptable solutions at $\theta=0,\pi$. 

Before ending this section it is worth noting another interesting limit of Eq.~(\ref{spectra}). For $m\gg n$ the BH mass is given by:

\begin{equation}
\label{spectra2}
M= \frac{m}{\sqrt{2}\mu}\;.
\end{equation}
This value for $M$ coincides with the upper limit of the band (\ref{Mband}). In this limit one also can attempt to find well behaved clouds 
as we did previously in the opposite limit $n\gg m$. The separation constants (\ref{Klmsubext}) become 
\begin{equation}
\label{Klmexp+}
K_{lm+} = l(l + 1) - \left(\frac{m}{2}\right)^2 + \sum_{k=1}^{\infty}c_{k}\left(\frac{m}{2}\right)^{2k}.
\end{equation}
On the other hand, in this limit the constants (\ref{Klm}) read
\begin{equation}
\label{Klm+}
K_{m+,\frac{1}{2}}^{\rm ext} = m^2 \;.
\end{equation}
The compatibility $K_{lm+}=K_{m+,\frac{1}{2}}^{\rm ext}$ between (\ref{Klmexp+}) and (\ref{Klm+}) is difficult to achieve in general because {\it a priori} 
it does not lead to a Diophantine equation like (\ref{Pell1}) due to the presence of the coefficients $c_k$ which are not integers. Nonetheless since we 
are assuming $m\gg n$, and according to (\ref{Klmexp+}) $\lambda=m/2$, there exist an asymptotic approximation 
to the separation constants (\ref{Klmexp+}) which is rather simple in the 
double limit $(m,\lambda)\rightarrow \infty$, in this case with $\lambda<m$ \cite{Hod2015b}. Remarkably, in this double limit
it is possible to find a quadratic Diophantine equation when imposing the above compatibility condition between the two kind of separation constants. 
Nonetheless, in this double limit we were unable to find solutions within the integers for the pair $(l,m)$ as we did in the opposite limit $m\ll n$. 
We thus conclude that  in the limit $m\gg n$ superregular cloud solution of the kind we are interested cannot be constructed 
(see the Appendix).

\subsection{The superregular extremal $R_{nlm}(r_{H}^{\rm ext})= 0$ ($\beta=3/2$) and more general scenarios with a Pell-Diophantine equation}
\label{sec:betathreehalf}

Besides the {\it superregular} scenario analyzed in the preceding section, one can consider an alternative situation where $R_{nlm}(r_{H}^{\rm ext})= 0$, 
but with bounded radial derivatives at the horizon some of which may vanish. According to Hod~\cite{Hod2012} this scenario is possible if one takes $\beta$ 
as a half integer with $\beta>1/2$. For simplicity we analyze only the case $\beta=3/2$, but for larger values the method is basically the same. 
According to (\ref{RHod2}), taking $\beta=3/2$ leads to radial functions where $R_{nlm}(r_{H}^{\rm ext})= 0$, $|R_{nlm}'(r_{H}^{\rm ext})|< \infty$ 
and $|R_{nlm}''(r_{H}^{\rm ext})|<\infty$. Now, using our method and proceeding in a similar fashion as for the case $\beta=1/2$, it is easy to see 
from Eq.~(\ref{CNEg1}) that assuming $R_{nlm}(r_{H}^{\rm ext})=0$ and $|R_{nlm}'(r_{H}^{\rm ext})|<\infty$, but $R_{nlm}'(r_{H}^{\rm ext})\neq 0$, 
one is led to the separation constants
\begin{equation}
\label{Klm2}
K_{m,\frac{3}{2}}^{\rm ext} = 2(1+m^2) - 2\mu^2M^2 \;,
\end{equation}
which, again, turn out to be different from (\ref{Klmsubext}).

Using these results and assumptions in Eq. (\ref{CNEg2}) we obtain,
\begin{equation}
\label{CNE3}
R''_{nlm}(r_{H}^{\rm ext}) =  \frac{\left(2\mu^2M^2- m^2\right)}{2M}R'_{nlm}(r_{H}^{\rm ext})\;.
\end{equation}

Unlike the scenario of previous section that has $R_{nlm}(r_{H}^{\rm ext})$ as free parameter, in this scenario 
$R'_{nlm}(r_{H}^{\rm ext})$ is the free parameter. We can choose for simplicity $R'_{nlm}(r_{H}^{\rm ext})=1$. 
Notice that the new kind of separation constants (\ref{Klm2}) correspond precisely to those obtained from (\ref{betaHod}) when $\beta=3/2$. 
Moreover for this value of $\beta$ one has from (\ref{kappasmooth0}) 
\begin{equation}
\label{kappasmooth2}
\kappa= 2 + n \;.
\end{equation}

\begin{figure}[h!]
\begin{center}
\includegraphics[width=0.45\textwidth]{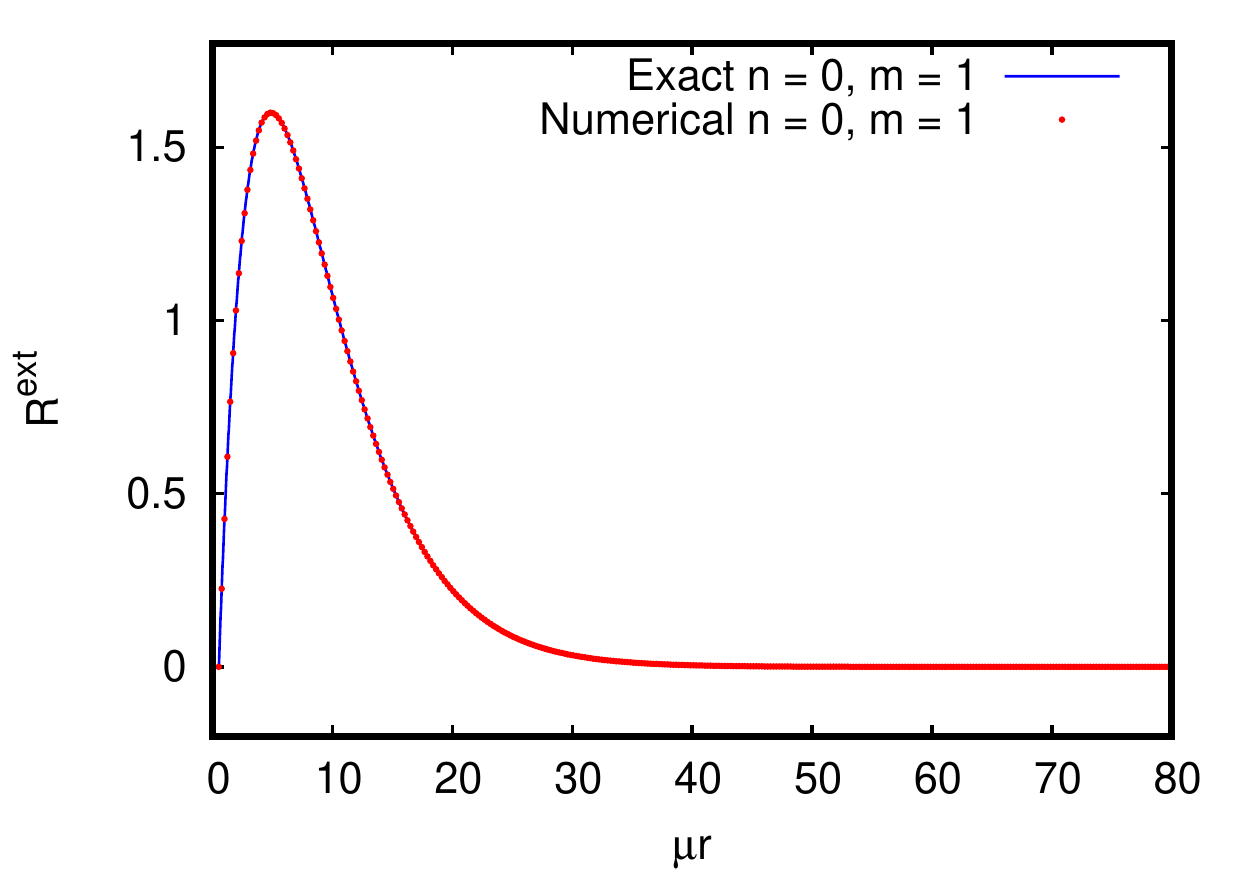}
\includegraphics[width=0.45\textwidth]{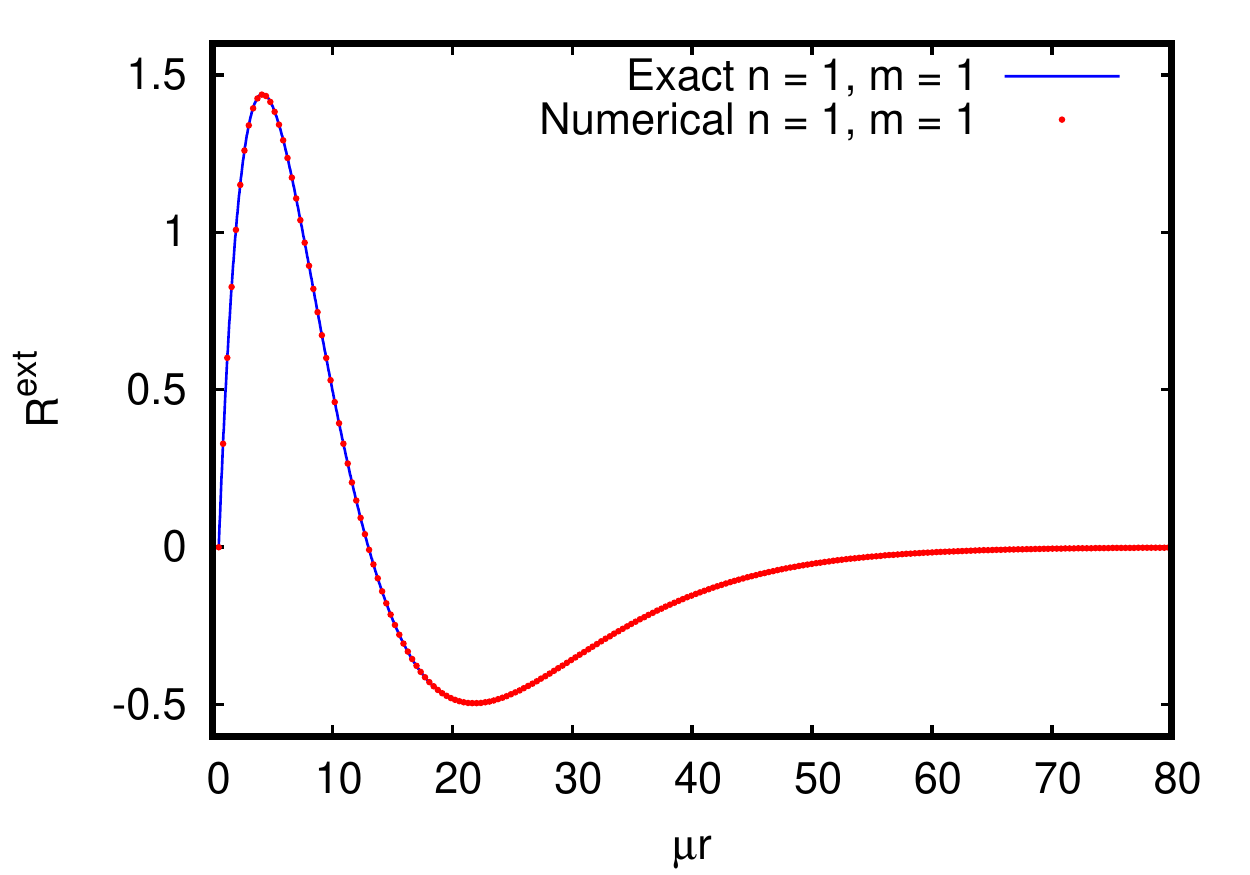}
\caption{Radial solutions $R^{\rm ext}$ associated with the extremal clouds with principal numbers $n = 0, 1$ and $m = 1$. 
The red dots correspond to the solutions obtained by solving numerically Eq. (\ref{radialE}) and the blue solid line corresponds to the exact 
solution given by Hod's Eq.~(\ref{RHod2}) with $\beta=3/2$. The constant $A$ is chosen so that $R^{'}_H=1$. Notice the agreement 
between both kind of solutions.}\label{fig:HodNumerical32}
\end{center}
\end{figure}

Given all these conditions one can compute the localized radial functions with the spectra (\ref{spectra1}) using (\ref{kappasmooth2}). Some solutions are shown in Figure~\ref{fig:HodNumerical32}. 

However, for this scenario we have again the same problem as before, i.e., 
the SH are not well behaved on the axis, except in the limit of large $n$ as in the previous section. 

This problem will still  be present for larger values of $\beta=k/2$ as in this case the separation constants as obtained from 
(\ref{betaHod}) or from the regularity conditions using the radial Teukolsky equation, will be different from (\ref{Klmsubext}). 
For instance, the case associated with $\beta=5/2$ is obtained as follows. From the regularity conditions Eq.~(\ref{CNEg1}) we see that we can take $R_{nlm}(r_{H}^{\rm ext})=0= R'_{nlm}(r_{H}^{\rm ext})$ 
without restricting the separation constants. Thus, using these conditions in Eq.~(\ref{CNEg2}), we conclude that $R''_{nlm}(r_{H}^{\rm ext}) \neq 0$ 
if the separation constants are $K_{m,\frac{5}{2}}^{\rm ext} = 2(3+ m^2) - 2\mu^2M^2$. These coincide with (\ref{betaHod}) for $\beta=5/2$. This 
scenario is more intricate numerically than the previous ones because we need to solve a differential equation involving $R'''_{nlm}(r)$ and take 
$R''_{nlm}(r_{H}^{\rm ext})$ as free parameter~\footnote{At this respect is worth stressing the following issue: if one solves the second order Teukolsky Eq.~(\ref{radialE}) with boundary conditions at the 
horizon given by $R_{nlm}(r_{H}^{\rm ext})=0= R'_{nlm}(r_{H}^{\rm ext})$ and $R''_{nlm}(r_{H}^{\rm ext})$ bounded 
one simply obtains the trivial vanishing solution. Nevertheless, with those 
two conditions imposed on the exact solution (\ref{RHod2}) for $\beta>5/2$, but excluding the values $\beta=k/2>5/2$, one 
does not obtain the trivial solution, but solutions where $R_{nlm}(r_{H}^{\rm ext})=0= R'_{nlm}(r_{H}^{\rm ext})$ but with derivatives higher than 
the second order unbounded at the horizon. The point is that, according to the uniqueness theorems of differential equations, 
given some ``initial'' data the uniqueness of the solution is warranted provided that smoothness on the derivatives is enforced ``initially''. 
To give a more simple example, let us consider the following differential equation: $y'(x)= \sqrt{x}$ subject to $y(0)=0$. 
One readily obtains the solution $y(x)= 2 x^{3/2}/3$. However, the trivial solution 
$y(x)\equiv 0$ is also perfectly valid. The point is that the trivial solution has all its derivatives vanishing, while the 
non trivial solution has higher derivatives unbounded at $x=0$. Thus, the uniqueness of this simple differential equation is not 
guaranteed given the value at $y(0)$ and with $y'(0)=0$ due the the lack of boundedness of the higher derivatives at $x=0$.}. But we shall not pursue the numerical analysis for $\beta>3/2$. 
Below we provide some generic results for any $\beta=k/2$ and complete our numerical results for $\beta=3/2$ in the limit $n\gg m$, 
for which the complete superregular cloud solution is well behaved. 

In general for any positive $\beta=k/2$ the separation constants (\ref{betaHod}) are given by 
\begin{eqnarray}
\label{Klmbeta}
K_{m,\beta}^{\rm ext} &=& \beta^2 -\frac{1}{4}+ 2(m^2-\mu^2 M^2) \nonumber \\
&=& (k^2-1)/4 + 2m^2 - 2\mu^2M^2 \nonumber \\
&=& j(j-1) + 2(m^2 -\mu^2M^2) \;.
\end{eqnarray}
where we used $k= 2j-1$,  with $j$ a positive integer. Then, with $\beta=k/2$ Eq.~(\ref{kappasmooth0}) 
leads to $\kappa= j+n$, and for this $\kappa$ the spectra (\ref{spectra1}) depend on the 
three integer numbers $n, m$ and $j$.

In the limit $n\gg m$ we have $M=m/2$, and (\ref{Klmbeta}) becomes
\begin{equation}
\label{Klmbetan}
K_{m,\beta}^{\rm ext} = \beta^2 -\frac{1}{4}+ \frac{3m^2}{2}= j(j-1) + \frac{3m^2}{2} \;.
\end{equation}
In this limit the separation constants (\ref{Klm2}) with $M=m/2$ are recovered from (\ref{Klmbetan}) taking  $\beta=3/2$, i.e. $j=2$.
For any $\beta$ we can make compatible (\ref{Klmbetan}) and (\ref{Klmexp}) if the pair $(l,m)$ satisfies 
\begin{eqnarray}
\label{mlninf2}
m &=& \sqrt{\frac{2}{3}\left[l(l + 1)- j(j-1)\right]} \nonumber \\
 &=& \sqrt{\frac{2}{3}\left[(l+\frac{1}{2})^2- \beta^2\right]} \nonumber \\
&=& \sqrt{\frac{2}{3}\left[(l+\frac{1}{2}+\beta)(l+\frac{1}{2}-\beta)\right]}
 \;.
\end{eqnarray}
This equation can be written exactly as Pell's Eq.~(\ref{Pell1}): 
\begin{equation}
\label{Pell2}
x^2_\beta - 6y^2_\beta = 1 \;,
\end{equation}
except that in this case,
\begin{equation}
x_\beta= \frac{2l+1}{2\beta}\;,\; y_\beta= \frac{m}{2\beta}\;.
\label{pairsxybeta}
\end{equation}
The trivial solution $(x_\beta,y_\beta)=(1,0)$ of Eq.~(\ref{Pell2}) 
corresponds to $m=0$, $l=(2\beta-1)/2=j-1$, whereas the fundamental solution 
$(x_{\beta,0},y_{\beta,0})=(5,2)$ corresponds to
\begin{eqnarray}
m &=& 4\beta= 2(2j-1)\;,\\
l &=& \frac{10\beta-1}{2}=5j-3 \;.
\end{eqnarray}
Thus, for $\beta=1/2$, i.e. $j=1$, we recover the results of previous section. 
For $\beta=3/2$, i.e. $j=2$, the fundamental solution $(5,2)$ of Eq.~(\ref{Pell2}) corresponds to
\begin{equation}
m= 6\;,\; l= 7 \;.
\end{equation}
Using the recurrence Eqs.~(\ref{Pellrecx}) and (\ref{Pellrecy}) we can generate all the subsequent pairs $(x_{\beta,n},y_{\beta,n})$, 
which are the same as for the value $\beta=1/2$, and then obtain $(l,m)$ from (\ref{pairsxybeta}).

\begin{figure}[h!]
\begin{center}
\includegraphics[width=0.45\textwidth]{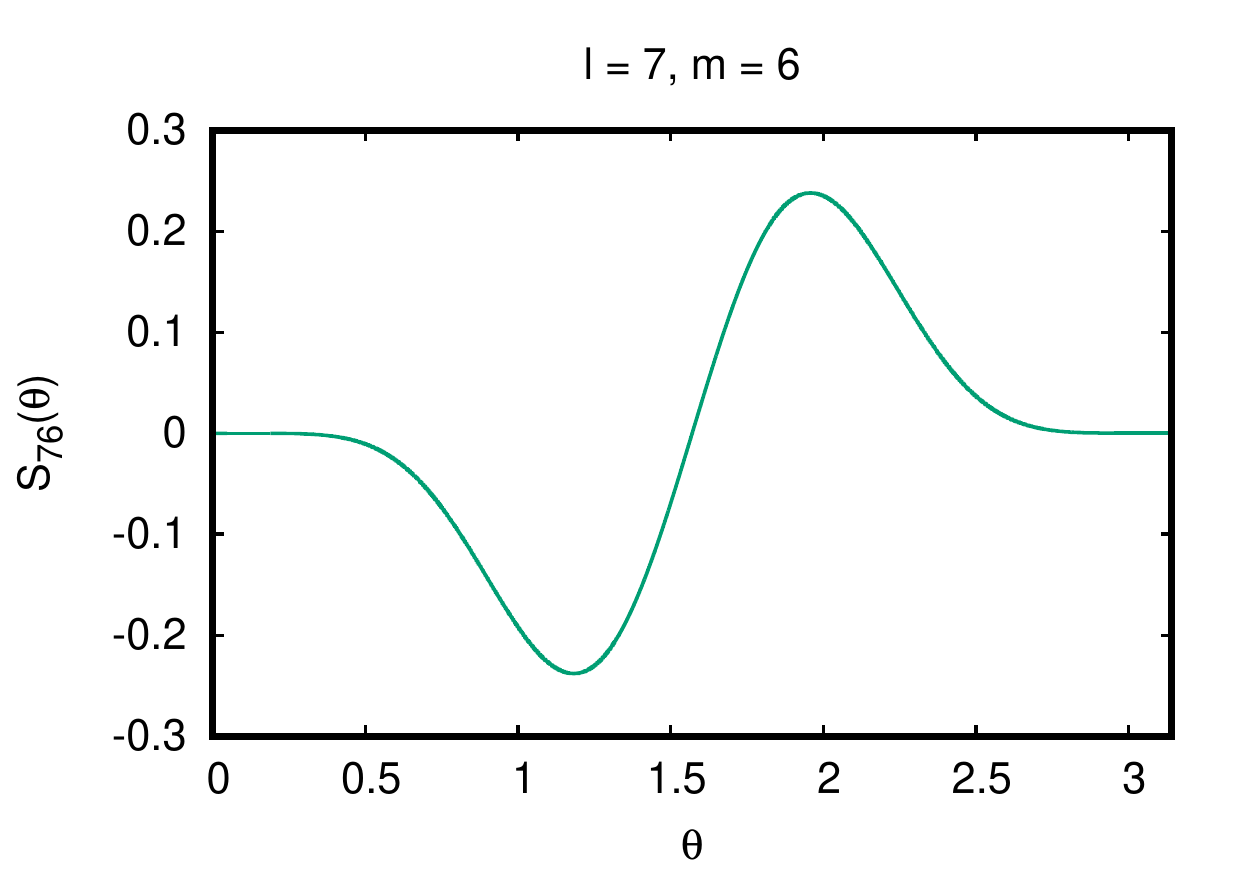}
\includegraphics[width=0.45\textwidth]{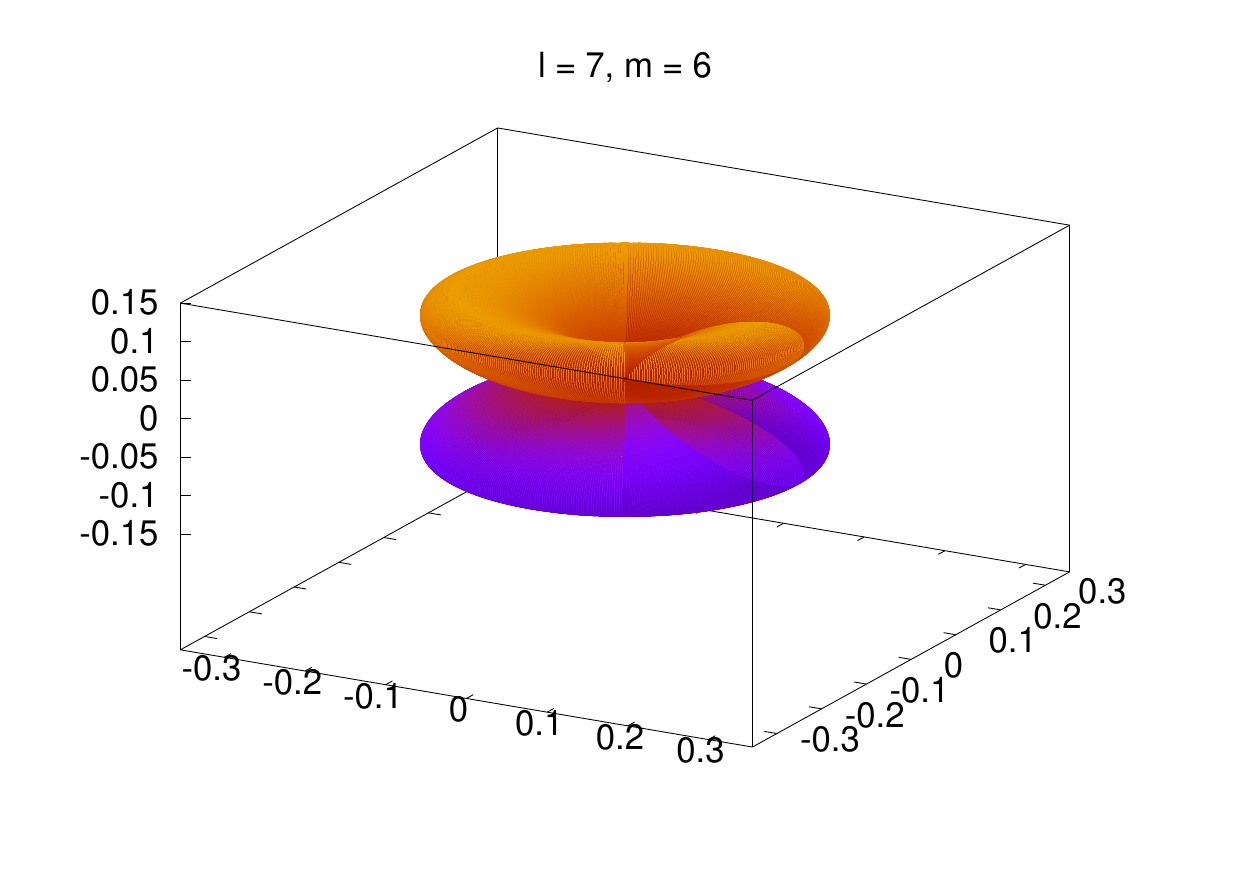}
\caption{Spheroidal harmonic (SH) $S_{lm}(\theta)$ (upper panel) and  $|S_{76}(\theta,\varphi)|$ (lower panel)
    for $m = 6$, $l = 7$ and $n = 10^9$ in the exact extremal case. The SH are well behaved on the axis of symmetry at $\theta=0,\pi$.}
\label{fig:Sm76}
\end{center}
\end{figure}

\begin{figure}[h!]
\begin{center}
\includegraphics[width=0.5\textwidth]{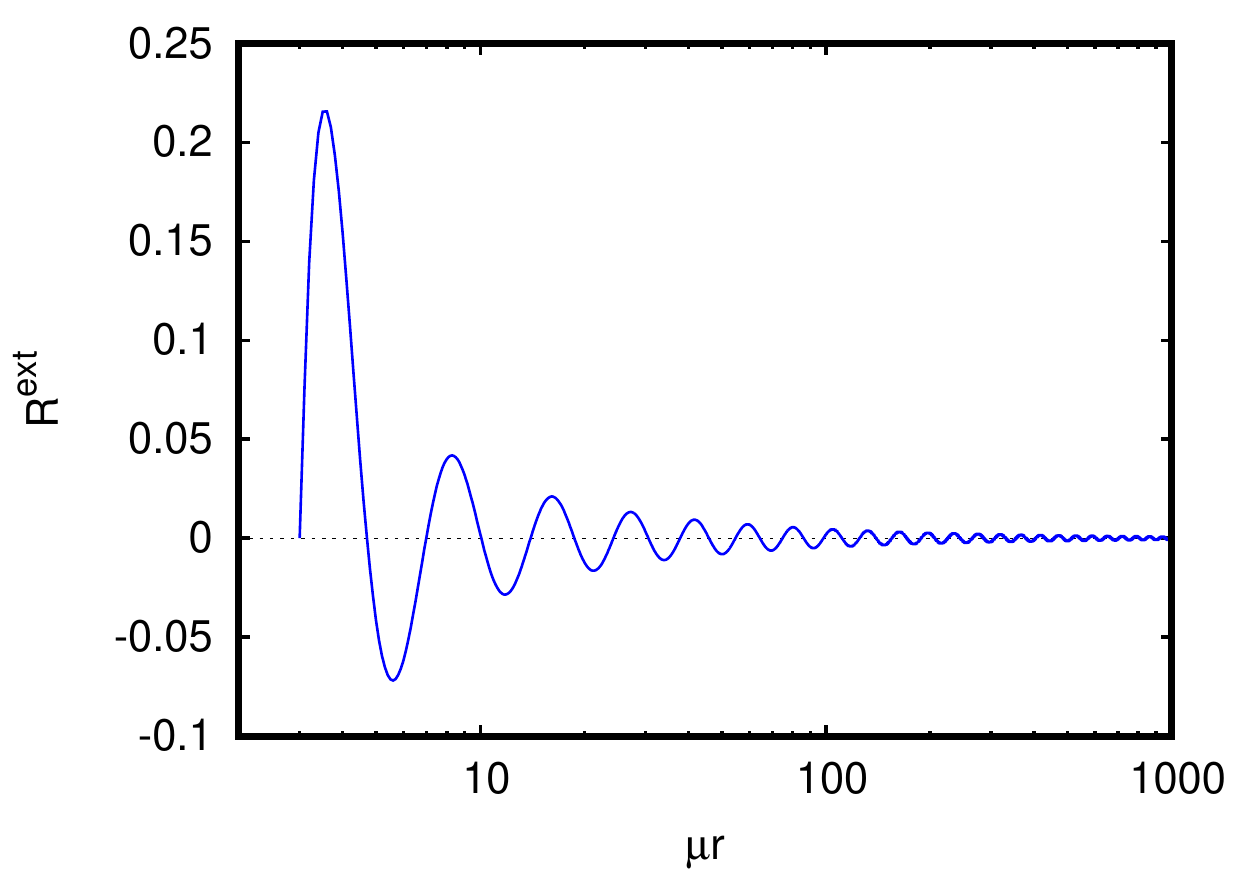}
  \caption{Radial solution $R^{\rm ext}$ with principal number $n = 10^9$ and $m = 6$ with 
$r_H^{\rm ext}=M=a$.} \label{fig:RInf2}
\end{center}
\end{figure}

Table~\ref{tab:Klmninf2} provides some examples of pairs $(l,m)$ satisfying the condition (\ref{mlninf2}) for $\beta=3/2$ which 
allows for the separation constants to verify $K_{lm}= K_{m,\frac{3}{2}}^{\rm ext}$. 
Table~\ref{tab:ninfinity2} shows the separation constants Eqs.~(\ref{Klm2}) and (\ref{Klmsubext}) for $m = 6$ and $l = 7$
  as $n$ increases. These constants were computed similarly as in Table~\ref{tab:ninfinity}. Figure~\ref{fig:Sm76}
depicts the SH, $S_{lm}(\theta)$ (upper panel) and 
$|S_{lm}(\theta,\varphi)|$ (lower panel) for $m = 6$ and $l = 7$ with $n = 10^9$ using $K_{m,\frac{3}{2}}^{\rm ext}$ given by Eq.~(\ref{Klm2}).
In this limit of large $n$ and for these values of $(l,m)$ for which $K_{lm}\approx K_{m,\frac{3}{2}}^{\rm ext}$ the SH are well behaved on the axis of symmetry. 
Figure~\ref{fig:RInf2} shows the radial function computed for $n = 10^9$ and $m=6$ which is associated 
with the SH in Figure~\ref{fig:Sm76}.

\begin{table}[htbp]
\begin{center}
\begin{tabular}{|c|c|c|}
\hline 
$l$ & $m$ & $K_{m,\frac{3}{2}}^{\rm ext} = K_{lm} $\\
\hline \hline 
1 & 0 & 1\\ \hline
7 & 6 & 56  \\ \hline                     
73 & 60 & 5402   \\ \hline
727 &  594  & 529256  \\ \hline
7201 & 5880   &  51861602 \\ \hline
71287 & 58206  &  5081907656\\ \hline
\end{tabular}
\caption{Columns 2 and 3 display a sample of integer values $l$ and $m$ satisfying the compatibility condition (\ref{mlninf2}) 
associated with the limit $n\rightarrow \infty$, and taking $\beta=3/2$. The pairs $(l,m)$ are obtained from (\ref{pairsxybeta}) 
using the numerical values $(x,y)$ of Table~\ref{tab:Klmninf}. The pair $l=1$, $m=0$ leads to trivial cloud solutions. Column 3 
shows the corresponding values of the separation constants (\ref{Klmbetan}) compatible with (\ref{Klmexp}).}
\label{tab:Klmninf2}
\end{center}
\end{table}

\begin{table}[htbp]
\begin{center}
\begin{tabular}{|c|c|c|}
\hline 
$n$ &  $K^{\rm ext}_{m,\frac{3}{2}}$  & $K_{lm}$ \\
\hline \hline 
10 &  55.586843519588086 &  54.996894379984859 \\ \hline
$10^2$ & 55.984455946957510 & 55.993599480311758 \\ \hline                     
$10^3$ & 55.999838648953897 & 55.999933561330181 \\ \hline
$10^4$ & 55.999998380648101 & 55.999999333208038 \\ \hline
$10^5$ & 55.999999983800649 & 55.999999993329681\\ \hline
$10^6$ & 55.999999999837996 & 55.999999999933294\\ \hline
$10^7$ & 55.999999999998380 & 55.999999999999332  \\ \hline
\end{tabular}
\caption{Behavior of separation constants (\ref{Klm2}) and (\ref{Klmsubext}) for $m = 6$, $l = 7$ as $n \rightarrow \infty$. 
These values are compatible with the exact value given in Table~\ref{tab:Klmninf2} for $m = 6$, $l = 7$.}
\label{tab:ninfinity2}
\end{center}
\end{table}

%%%%%%%%%%%%%%%%%%%%%%%%%%%%%%%%%%%%%%%%%%%%%%%%%%%%%%%%%%%%%%%%%%%%%%
%%%%%%%%%%%%%%% Section: Conclusion %%%%%%%%%%%%%%%%%%%%%%%%%%%%%%%%%%
%%%%%%%%%%%%%%%%%%%%%%%%%%%%%%%%%%%%%%%%%%%%%%%%%%%%%%%%%%%%%%%%%%%%%%
\bigskip

\section{Conclusions}
\label{sec:conclusion}
Contrary to our initial expectations that consisted in finding regular boson clouds  in the background of extremal Kerr BH with bounded 
field and its derivatives at the horizon, i.e., {\it superregular} clouds 
(as opposed to the corresponding clouds with only the field, but not its derivatives 
bounded at the horizon) we found instead that these {\it superregular} clouds do not exist in general when using Boyer-Lindquist coordinates. 
Despite the fact that
we computed perfectly valid radial functions with bounded derivatives at the horizon by finding in addition their exact spectrum
analytically, we observed, however, that the angular functions, which were supposed to be provided by the spheroidal harmonics (SH),
do not actually ``exist'' for $m\neq 0$ (i.e. they are unbounded at $\theta=0,\pi$). Such an unacceptable behavior 
is due to an inconsistency between the separation constants obtained from regularity in the SH on the axis of symmetry and the 
separation constants obtained from regularity conditions of the radial function at the horizon. Intriguingly this inconsistency 
does not happen in the subextremal case.  Nonetheless, in the extremal case we found that in the limit of a large number of 
nodes for the radial function ($n\rightarrow \infty$) it is possible to avoid the inconsistency and find superregular clouds provided that the 
numbers $(l,m)$ obeys a quadratic Diophantine equation of Pell's type. This result establishes an interesting connection 
between field theory in curved spacetimes with number theory. At the same time, this conclusion indicates 
that acceptable cloud solutions with more general pairs $(l,m)$ (with $n$ not necessarily large) fixes a value $\beta$ given by (\ref{betaHod}) that excludes 
the half-positive integer values. Otherwise, the inconsistency reemerges. Thus, in the 
exact solution (\ref{RHod2}), $\beta$ is {\it not} a free parameter, but a number provided by (\ref{betaHod}), which 
contain quantities that are involved in both the radial and the angular equations. 

All these arguments indicate that in the extremal case the boundedness conditions imposed 
on the radial derivatives using the usual $r$ coordinate are too restrictive and might not be necessary. For instance, the analytic solution (\ref{RHod2}) shows that 
near the horizon $R'(r)\sim {\rm const} \times (\beta-1/2) z^{\beta - 3/2} L_n^{(2\beta)}(z)$ so for $1/2< \beta < 3/2$, $R'$ blows up at the horizon. 
However, what is important physically is, for instance, the scalar (covariant) quantities formed by the full kinetic term. 
In this term it appears the quantity $g^{rr} R'^2\sim {\rm const} \times  z^{2\beta - 1} (L_n^{(2\beta)}(z)/\rho(r,\theta))^2$ which is finite at the horizon 
despite the fact that $R'$ is not. This suggests the use of a Wheeler's {\it tortoise} type of coordinate 
defined as $dr^{*}= (r^2+M^2)/(r-M)^2 dr$ (or alternatively a proper radial coordinate $dl= dr/(r-M)$). In this case ${\bar R}'(r^*)= R' dr/dr^*$, where ${\bar R}(r^*)= R(r)$, 
and near the horizon ${\bar R}'^2 \sim {\rm const} \times (R')^2 (r-M)^4 \sim  
{\rm const} \times z^{2\beta + 1} (L_n^{(2\beta)})^2$ which behaves exactly as $R'^2 z^4$ near the horizon. 
Therefore, even if $R'$ blows up at the horizon in the original 
$r$ coordinate as $R'\sim 1/(r-M)^\epsilon \sim 1/z^\epsilon$ with $0<\epsilon<1$ 
in the range $1/2 <\beta < 3/2$, ${\bar R}'$ not only remains finite, but vanishes, and for larger $\beta$, ${\bar R}'$ also vanishes. 
The same conclusion holds for higher derivatives. 
In view of this we conclude that using the coordinate $r^*$ it may be possible to find superregular clouds numerically 
for any number of nodes $n$ and for any pair of angular numbers $(l,m)$ with $ |m| \leq l$. It is however not clear in which place 
of the domain of outer communication of the BH one should give adequate boundary conditions to start the numerical integration in view that 
the coordinate $r^*$ covers the domain $-\infty< r^*<\infty$, i.e. from the horizon at $r^*=-\infty$ to spatial infinite $r^*=+\infty$.

In a forthcoming paper~\cite{Garcia2019b}, we plan to report a similar analysis for clouds in the background of Kerr-Newman spacetime 
(i.e. one where the BH is rotating and charged) where we will consider the subextremal, near extremal and exact (superregular) 
scenarios and try to implement the coordinate $r^*$ when needed 
and compare our numerical results with \cite{Hod2015,Degollado2013,Benone2014}.

%%%%%%%%%%%%%%%%%%%%%%%%%%%%%%%%%%%%%%%%%%%%%%%%%%%%%%%%%%%%%%%%%%%%
%%%%%%%%%%%%%%%%%%%%    Acknowledgments     %%%%%%%%%%%%%%%%%%%%%%%%
%%%%%%%%%%%%%%%%%%%%%%%%%%%%%%%%%%%%%%%%%%%%%%%%%%%%%%%%%%%%%%%%%%%%

\acknowledgments 
This work was supported partially by DGAPA--UNAM grants IN111719 and SEP--CONACYT grants CB--166656. G.G. acknowledges CONACYT scholarship 
291036. We thank C. Herdeiro, S. Hod and J.C. Degollado for valuable suggestions and discussions.
\bigskip

%%%%%%%%%%%%%%%%%%%%%%%%%%%%%%%%%%%%%%%%%%%%%%%%%%%%%%%%%%%%%%%%%%%%
%%%%%%%%%%%%%%%%%%%%    Appendix            %%%%%%%%%%%%%%%%%%%%%%%%
%%%%%%%%%%%%%%%%%%%%%%%%%%%%%%%%%%%%%%%%%%%%%%%%%%%%%%%%%%%%%%%%%%%%

\section{Appendix} 
\label{sec: appendix}

In the following we analyze in more detail the limit $m\gg n$ mentioned at the end of Sec.~\ref{sec:betaonehalf}. 
Since in this limit the quantity 
$\lambda:= a^2(\mu^2 - \omega^2)$ introduced in Sec.~\ref{sec:numerics} is $\lambda=m/2$, then $\lambda$ is also large 
in this limit, and thus we can use the {\it double limit} asymptotic analysis for the separation constants $K_{lm}$ provided by 
Hod~\cite{Hod2015b}. In order to do so, and 
to avoid any possible confusions about the notation, it is convenient to consider the differential equation for the SH 
as written in~\cite{Hod2015b}:

\begin{eqnarray}
\label{Hodangular}
&& \frac{1}{\sin\theta}\frac{\partial}{\partial\theta}\left(\sin\theta\frac{\partial S_{lm}(\theta)}{\partial\theta}\right) \nonumber\\
&& + \left[A_{lm} + c^{2}\cos^2\theta - \frac{m^{2}}{\sin^{2}\theta}\right]S_{lm}(\theta) = 0 \;.
\end{eqnarray}
Comparing the angular Eq.~(\ref{angularE}) with Eq.~(\ref{Hodangular}) we obtain the relationship between our notation with Hod's 
\cite{Hod2015b}.
\begin{eqnarray}
\label{KAlm}
\label{cHod}
c^{2}  &=& -\lambda = -M^{2}\left(\mu^{2} - \omega^{2}\right) \;,\\
K_{lm} &=& A_{lm} + c^{2} \;.
\end{eqnarray}
According to~\cite{Hod2015b}, when $\left\lbrace\mid c\mid, m\right\rbrace \rightarrow \infty$ with $m^{2} > c^{2}$ and $m/c$ a finite number, 
the separation constants $A_{lm}$ become
\begin{eqnarray}
\label{AlmHod}
A_{lm} &\approx & m^{2} + \left[2N + 1\right]\sqrt{m^{2} - c^{2}} \;, \\
N &=& l-|m| \;,
\end{eqnarray}  
where $N$ is a non-negative integer. In this paper we focused only in non-negative $m$, thus $N=l-m$. 
From Eq.~(\ref{spectra}) we found first $\mu M = m/\sqrt{2}$ when $m \gg n$ and from Eqs.~(\ref{fluxcond2}) and (\ref{OmH})
$\omega= \mu/\sqrt{2}$. Then we obtained the separation constants $K_{lm+}$ (\ref{Klm+}). 
From Eq.~(\ref{cHod}), and in this large $m$ limit we obtain $c^{2} = -m^{2}/4$. We see then that the three conditions to 
apply the asymptotic formula (\ref{AlmHod}) are met: $m^2>c^2$, $|c|$ is large in this limit, and finally $m/|c|=2$.  
Thus, from Eqs.~(\ref{KAlm})--(\ref{AlmHod}) we find $K_{lm}$ for $m$ large:
\begin{equation}
\label{KlmHod}
K_{lm+}= \frac{3m^2}{4} + \left[2(l-m) + 1\right] \frac{m}{2}\sqrt{5} \,.
\end{equation}
Consistency between the radial and angular parts of the solution require 
$K_{m+,\frac{1}{2}}^{\rm ext} = K_{lm+}$, as given by Eqs.~(\ref{Klm+}) and Eq.~(\ref{KlmHod}). 
For instance if $l=m$ and for $m$ sufficiently large $K_{lm+} \approx 3m^2/4$, and the consistency condition leads to 
$m^2=3m^2/4$ which is impossible to achieve for large $m$. Now, when taking into account the complete $K_{lm+}$ given by 
(\ref{KlmHod}), the consistency condition leads to the following 
quadratic Diophantine equation for $(l,m)$:
\begin{equation}
\label{quadratic1}
al^{2} + blm + cm^{2} + dl + em + f = 0 \;,
\end{equation}
where $a = 80$, $b = -160$, $c = 79$, $d = 80$, $e = -80$ and $f = 20$. This equation can be written as
\begin{equation}
al^{2} + \left(bm + d\right)l + \left(cm^{2} + em + f\right) = 0 \;,  
\end{equation}
this is a second degree equation for $l$. A necessary condition for this equation to have integer solutions is that the discriminant is a 
perfect square:
\begin{equation}
(bm + d)^{2} - 4a(cm^{2} + em + f) = u^{2} \;,
\end{equation}
where $u$ is a integer number. The last equation can be rewritten as
\begin{equation}
\label{quadm}
pm^{2} + qm + r - u^{2} = 0 \;,
\end{equation}
where $p = b^{2} - 4ac$, $q = 2bd - 4ae$ and $r = d^{2} - 4af$. Equation (\ref{quadm}) is a second degree equation in $m$ which 
may have integer solutions if the discriminant is a perfect square:
\begin{equation}
q^{2} - 4p\left(r - u^{2}\right) = v^{2} \;,    
\end{equation}
where $v$ is an integer number. This equation can be written as $v^{2} -Du^{2} = t$, with $D = 4p$ and $t = q^{2} - 4pr$. For our particular case we get
$p = 320$, $q = 0$, $D = 1280$, $r = 0$ and $t = 0$. So the Diophantine equation to solve is $v^{2} - 1280u^{2} = 0$, which does not admit integer solutions 
for large $m$ (i.e. $v\neq 0$ and $u\neq 0$). We conclude that for $m \gg n$ there are no consistent cloud solutions contrary to what happens in the opposite limit $m \ll n$.

%%%%%%%%%%%%%%%%%%%%%%%%%%%%%%%%%%%%%%%%%%%%%%%%%%%%%%%%%%%%%%%%%%%%
%%%%%%%%%%%%%%%%%%%%%%%       BIBLIOGRAPHY     %%%%%%%%%%%%%%%%%%%%%
%%%%%%%%%%%%%%%%%%%%%%%%%%%%%%%%%%%%%%%%%%%%%%%%%%%%%%%%%%%%%%%%%%%%

\bibliography{referencias}{}

\end{document}